\documentclass[pdflatex,sn-mathphys-num]{sn-jnl}

\usepackage{graphicx}%
\usepackage{multirow}%
\usepackage{amsmath,amssymb,amsfonts}%
\usepackage{amsthm}%
\usepackage{mathrsfs}%
\usepackage[title]{appendix}%
\usepackage{xcolor}%
\usepackage{textcomp}%
\usepackage{manyfoot}%
\usepackage{booktabs}%
\usepackage{algorithm}%
\usepackage{algorithmicx}%
\usepackage{algpseudocode}%
\usepackage{listings}%
\usepackage{subfigure}

\usepackage{array}        
\usepackage{makecell}     
\usepackage{rotating}     

\usepackage{amssymb}  
\usepackage{mathrsfs} 

\theoremstyle{thmstyleone}%
\theoremstyle{thmstyletwo}%

\theoremstyle{thmstylethree}%

\begin{document}

\title[Article Title]{Deep Joint Source-Channel Coding for Wireless Video Transmission with Asymmetric Context}


\author[1]{\fnm{Xuechen} \sur{Chen}}\email{chenxuec@csu.edu.cn}

\author[2]{\fnm{Junting} \sur{Li}}\email{234712145@csu.edu.cn}

\author[3]{\fnm{Chuang} \sur{Chen}}\email{chenchuang34@huawei.com}

\author[1]{\fnm{Hairong} \sur{Lin}}\email{haironglin@csu.edu.cn}

\author*[1]{\fnm{Yishen} \sur{Li}}\email{liyishen016@gmail.com}

\affil*[1]{\orgdiv{School of Electronic Information}, \orgname{Central South University}, \orgaddress{ \city{Changsha}, \postcode{410083}, \state{Hunan}, \country{China}}}

\affil[2]{\orgdiv{School of Computer Science and Technology}, \orgname{Central South University}, \orgaddress{ \city{Changsha}, \postcode{410083}, \state{Hunan}, \country{China}}}

\affil[3]{\orgdiv{2012 Lab}, \orgname{Huawei}, \orgaddress{ \city{Shenzhen}, \postcode{518000}, \state{Guangdong}, \country{China}}}


\abstract{In this paper, we propose a high-efficiency deep joint source-channel coding (JSCC) method for video transmission based on conditional coding with asymmetric context. The conditional coding-based neural video compression requires to predict the encoding and decoding conditions from the same context which includes the same reconstructed frames. However in JSCC schemes which fall into pseudo-analog transmission, the encoder cannot infer the same reconstructed frames as the decoder even a pipeline of the simulated transmission is constructed at the encoder. In the proposed method, without such a pipeline, we guide and design neural networks to learn encoding and decoding conditions from asymmetric contexts. Additionally, we introduce feature propagation, which allows intermediate features to be independently propagated at the encoder and decoder and help to generate conditions, enabling the framework to greatly leverage temporal correlation while mitigating the problem of error accumulation. To further exploit the performance of the proposed transmission framework, we implement content-adaptive coding which achieves variable bandwidth transmission using entropy models and masking mechanisms. Experimental results demonstrate that our method outperforms existing deep video transmission frameworks in terms of performance and effectively mitigates the error accumulation. By mitigating the error accumulation, our schemes can reduce the frequency of inserting intra-frame coding modes, further enhancing performance.}




\maketitle

\section{Introduction}\label{sec1}
Video content accounts for more than 80 percent of online traffic, and this percentage has a tendency to increase gradually \cite{cisco}. The wireless video transmission task is divided into two subproblems: the source encoder compresses the video into the shortest possible bit stream while maintaining comparable visual effects, and the channel encoder adds redundancy to the compressed bit stream to resist the effects of channel interference. The video transmission system is thus modular in design. Separate transmission systems are easy to deploy in various scenarios and applications due to their independent design and optimization of individual modules and have achieved excellent results. However, with the rise of new video applications such as virtual reality (VR), mobile video services and etc., the limitations of the separate design have become more prominent. This design criterion loses its optimality due to the energy constraints of the devices and the time-varying nature of the channel quality. Moreover, the separate system is prone to the ``cliff effect"-- a drastic drop in performance when the channel capacity is lower than the communication rate. Most of the systems use the automatic repeat request (ARQ) mechanism and adaptive bit rate (ABR) strategy to adjust the bit rate according to the real-time channel state, which greatly affects the coding and communication efficiency.

In order to solve these problems, the joint design of source and channel coding for communication systems has been widely studied \cite{jscc1,jscc2}. It has been theoretically demonstrated that under finite delay constraints, over a limited range of block lengths, JSCC has lower distortion performance than independently designed schemes \cite{jscc3,jscc4,jscc5,jscc6}. Early JSCC systems, directly followed the modules of the separate system, but jointly optimized the individual modules that is an intuitive idea. However, such system does not provide sufficient performance gain. A more common and higher-performance approach is to not use any intermediate digital modules that involve bit operation and to transmit in a pseudo-analog way, e.g., using amplitude modulation (AM) or high-precision QAM (64K-QAM). In this relatively simple design, the video source is mapped directly to the channel symbols, and the ``cliff effect" is eliminated. From an information-theoretic point of view, by transmitting independent Gaussian sources in an additional white Gaussian noise (AWGN) channel, the uncoded coding is proved to be theoretically up to the Shannon boundary. Thus based on analogue transmission, JSCC exhibits lower computational complexity and good performance.

However, hand-designed methods and simple uncoded transmissions do not perform well on complex sources, especially when the performance is further degraded by bandwidth mismatch. Recently, deep learning-based JSCC methods have attracted great interest in wireless communications. It enables JSCC to achieve efficient transmission from a semantic point of view via artificial neural networks (ANNs). Intuitively, training an auto-encoder as a function of mapping sources to channel symbols solves the complexity problem of manually designing efficient JSCC systems. 
The current Deep JSCC schemes in \cite{djscc1,djscc2,djscc3,djscc4,djscc5,djscc6,djscc7,djscc8,djscc9,djscc10} have been shown to outperform separate schemes based on the classical image compression methods such as JPEG and the channel coding such as LDPC. However little research has been done on Deep JSCC for video transmission, where there remains extreme similarity between video frames. For cutting-edge video services, the frame rate can be up to 32fps and 64fps, which means that this similarity may last for hundreds of frames. Therefore, for video signals, it is not efficient to directly follow the Deep JSCC system designed for images. 

Recently Deep Learning (DL)-based video compression methods have been investigated enthusiastically. Most of them inherit the predictive coding framework from the traditional work to encode the residual and overlook the capabilities of DL which can automatically explore correlations in a huge space. Specifically, they assume that the current pixel only exists the correlation with the predicted pixel. In contrast, the Deep Contextual Video Coding (DCVC) \cite{dcvc} and its successor DCVC with diverse contexts (DCVC-DC) \cite{dcvc-dc} use already reconstructed frames as context to learn conditions which are defined as learnable contextual features with arbitrary dimensions. The extracted conditions are fed into the conditional coding network along with the current frame. In this way, taking into account that one pixel in the current frame correlates to all the pixels in the previously decoded frames and the pixels already been decoded in the current frame, the time-domain correlations are implicitly learned. Hence the higher video quality can be rendered. 

In \cite{dcvc} and \cite{dcvc-dc}, the reconstructed frames need to be obtained at the encoder. This is feasible if only considering video compression. However, as for Deep JSCC of video \cite{dvst}, due to the nature of analog coding, the encoder cannot infer the same reconstructed frames and motion vectors with the ones at the decoder even though a pipeline of the simulated transmission is constructed at the encoder. It is inappropriate to directly use the network structure of DCVC as the feature extraction part for Deep JSCC of video.

To solve above problems, we propose a video transmission scheme with asymmetric diverse context which includes video frames, intermediate features, and motion information, where the encoder and decoder separately predict conditions to learn inter-frame correlations from different contexts. Specifically, the encoder directly employs the original frame as part of the context to generate the coding conditions, while at the decoder the reconstructed frame is used accordingly. We first acquire motion information of neighboring ground-truth frames at the encoder and transmit it to the decoder. 
Moreover, we retain the intermediate features at the encoder and decoder, respectively, and propagate them to the codec for the next frame along with the motion information. 
This combination of feature propagation and asymmetric context provides multiple frame reference patterns for the encoder to encode the current frame, which can help exploit long-range time-domain correlations.

After the condition encoder provides a compact latent representation of the video source, we use an entropy model to predict the entropy value of this latent variable, which indicates the amount of information carried by the latent variable. Later, the entropy of latent variables is inputted into a policy network resulting in a mask, with the amount of `1' in the mask indicating the predicted bandwidth required for transmission. Lastly, an entropy coding network is used to adjust the information embedded in each channel according to the mask. In this way, we can achieve variable bandwidth transmission depending on the video content.
The contributions of the proposed method can be summarized as follows:
\begin{itemize}
\item We propose a novel end-to-end deep learning video transmission framework, where we analyze the discrepancy between the conditions in JSCC and those in compressed coding, and propose to learn coding by neural networks based on asymmetric contexts.
\item We introduce feature propagation into asymmetric conditional coding, where a set of features is propagated separately by the encoder and decoder and the feature reconstruction term is included in the loss function. Those features implicitly describe the correlation among a long range of frames, providing a reference pattern related to multiple frames for coding the current frame.
\item Considering the excellent performance of hybrid entropy model in video compression, we propose content-adaptive transmission guided by entropy model. We use the entropy of latent variables as a criterion for allocating bandwidths, which are adaptively adjusted by a policy network and an entropy coding network.
\end{itemize}

\section{RELATED WORKS}
\subsection{Residual Coding-based and Conditional Coding based-Neural Video Compression}
The pioneer work of Neural Video Compression (NVC) in \cite{dvc6} follows the framework of traditional video coding also known as residual coding-based framework, which uses an optical flow network and warp instead of the original motion coding to generate prediction frames, and then the residual between the predicted frame and the current frame is encoded by a hyper prior-based network. Many subsequent works \cite{dvc1,dvc3,dvc4,dvc5} also inherit the residual coding-based framework. For example, \cite{dvc1} performs motion coding and residual coding in the feature space, which reduces the computational cost. In \cite{dvc3}, the authors combined the residual coding with multi-scale motion estimation. Under the residual coding frameworks, Lin \emph{et al.} proposed to decompose the motion modeling \cite{dvc4} and Huang \emph{et al.} proposed a learned semantic representation for machine-friendly video compression \cite{dvc5}.

Residual coding only employs subtraction to remove redundancy, and it can only explicitly utilize the similarity of pixels in adjacent frames, at the same location. In contrast, conditional coding provides more exploitable space and more flexible design, since the condition can be any information that helps to improve the coding performance. Generally speaking, any prior information that can be used to improve the coding efficiency can be regarded as a condition, such as the previous video frame, motion information, or the semantic expression of the previous frame. In \cite{dcvc}, the authors define the condition as learnable contextual features with arbitrary dimensions. 
Following such a conditional coding framework, DCVC-HEM \cite{DCVC-HEM} designed efficient hybrid prior models by utilizing both spatial and temporal contexts. DCVC-TCM \cite{DCVC-TCM} utilized the propagation of information in the feature domain to improve the coding efficiency. DCVC-DC \cite{dcvc-dc} proposed that increasing the context diversity can improve the coding efficiency while the contexts are complementary to each other and have larger chance to provide good reference for reducing redundancy. 

The residual coding and the conditional coding are used to mine the inter-frame correlation of the video. Both residual coding-based and conditional coding-based NVC require further use of entropy coding to eliminate the redundancy in the latent variable. Therefore, another hotspot in NVC research is entropy modeling, which predicts the density of the latent variable for lossless entropy coding. 

However, entropy coding is discarded by most works focusing on JSCC due to the fact that it involves bit operations, and the entropy model was consequently discarded. On the contrary, in this paper, we introduce the efficient entropy model for content-adaptive coding. Through the network, variable bandwidth transmission can be achieved with the masking mechanism. 

\subsection{JSCC for Video Transmission}
The JSCC for video transmission has been of continuous interest to researchers for many years, with early work focusing not on codec design but mainly on rate-distortion optimization for JSCC \cite{jsccv1,jsccv2,jsccv3,jsccv4}. There are also a large amount of works based on Scalable Video Coding (SVC) that jointly optimizes source coding and channel coding \cite{jsccv6}. SoftCast \cite{soft1,soft2} is a radically innovative work that introduces pseudo-analog coding into video transmission with a series of linear transformations instead of the original entropy coding, channel coding to map the video from the pixel space to channel symbols. Since the channel symbols are linearly related to the video pixels, the quality of the video reconstruction is also linearly related to the channel quality, providing a ``one-size-fits-all" feature that can be adapted to different channel environments. Afterwards, Lan \emph{et al.} \cite{soft3}, Gui \emph{et al.} \cite{soft4} and He \emph{et al.} \cite{soft5} proposed to combine such pseudo-analog coding with existing digital coding systems, compressed sensing and non-orthogonal frequency division multiple access, respectively. The recent performance of DL in image transmission triggered the research on the DL-based video transmission problem. DeepWive proposed in \cite{deepwive} directly encodes each frame using the network structure of the deep JSCC system for image transmission. In predictive coding, it relies on reference frames before and after to capture correlations between frames using a scale-space flow model. 
However, directly adopting the deep JSCC system designed for image transmission shows inefficiency with respect to the complexity of the video source. 
Subsequently, Deep Video Semantic Transmission (DVST) \cite{dvst} uses a nonlinear transform and conditional coding architecture to extract semantic features. In DVST \cite{dvst}, a pipeline of simulated transmission at the encoder is set up to obtain the reconstructed previous frame and reconstructed motion information. 
This strategy is intuitive to capture consistent contexts at the encoder and decoder like DCVC \cite{dcvc}. However, contextual asymmetry is an important feature of pseudo-analog video transmission as the simulated reconstructed frames at the encoder are not exactly the same with the ones recovered by the decoder due to the randomness during the transmission. As a matter of fact, DNNs have the ability to cope with situations when the encoder and decoder have different contexts. Hence in this paper, we recognize the phenomenon and guide the network to learn from asymmetric contexts to effectively mitigate the errors as well as the accumulation of errors. Specifically, different from \cite{deepwive}, we adopt conditional coding networks and use the previous frames only as the reference for low-delay consideration. Moreover, unlike DVST \cite{dvst}, we directly use the previous ground-truth information and motion information as a context at the encoder. Then the codec learning from asymmetric contexts is designed. The details would be given in the next section.

Besides, compared to DVST \cite{dvst}, we introduce feature propagation into asymmetric conditional coding. Although the coding of the current frame also utilizes information from multiple previous frames in DVST \cite{dvst}, it did not adopt different weighting parameters $w_t$ for different P-frame codecs, as implemented in DCVC-DC \cite{dcvc-dc}. Until the work of DCVC-DC \cite{dcvc-dc}, feature propagation was first proposed and it can alleviate the problem of error propagation. In this paper, in addition to setting different $w_t$, we have introduced the feature reconstruction term in our loss function to ensure the generation of high-quality features, which can further alleviate the problem of error propagation. Additionally, in DVST \cite{dvst}, intermediate features assist in coding of results after entropy model estimation. Our method indirectly inputs the condition information generated by the intermediate features into the entropy model, which can effectively use the information of multiple frames to assist the entropy model parameter estimation.

In addition, adaptive bandwidth transmission is a key issue in wireless video communication. In the traditional separated video transmission framework, the video can be encoded and compressed into bit streams of different dimensions depending on the video content. In the recent Deep JSCC framework, this problem is also noted. Reference \cite{djscc4} introduces a policy network to dynamically adjust channel bandwidth based on channel conditions and image content. Compared to it, we make use of the entropy model to achieve adaptive rate transmission. In \cite{jsccv8}, a decision module is introduced to consider Channel State Information (CSI) and specific input signals to infer required channel bandwidth, which does not consider the whole content. In \cite{jsccv9}, a deep joint source-channel coding model for image protection and deprotection is proposed while the fixed channel bandwidth is used. In \cite{jsccv10}, a Signal-to-Noise Ratio (SNR) adaptive mechanism is adopted to deal with the wireless channel variations. Specifically, it inputs the SNR into an attention-based feature module to adjust bandwidth. Compared with it, we implement the adaptive control of the required channel bandwidth based on the video content. In this paper, we use the entropy model to get entropy and calculate the required bandwidth through the policy network. DVST \cite{dvst} needs to construct dynamic neural networks for each channel to achieve variable-rate coding and ours only need the policy network to achieve the bandwidth allocation, which can reduce the size of model scale.

In this paper, we link the information entropy and the rate in the JSCC framework to achieve more efficient adaptive rate transmission by introducing the entropy model that has achieved great success in the traditional compression field. We achieve adaptive bandwidth based on the video content, which can save transmission bandwidth while maintaining performance.

\begin{figure*}[!t]
\center
\includegraphics[scale = 0.4]{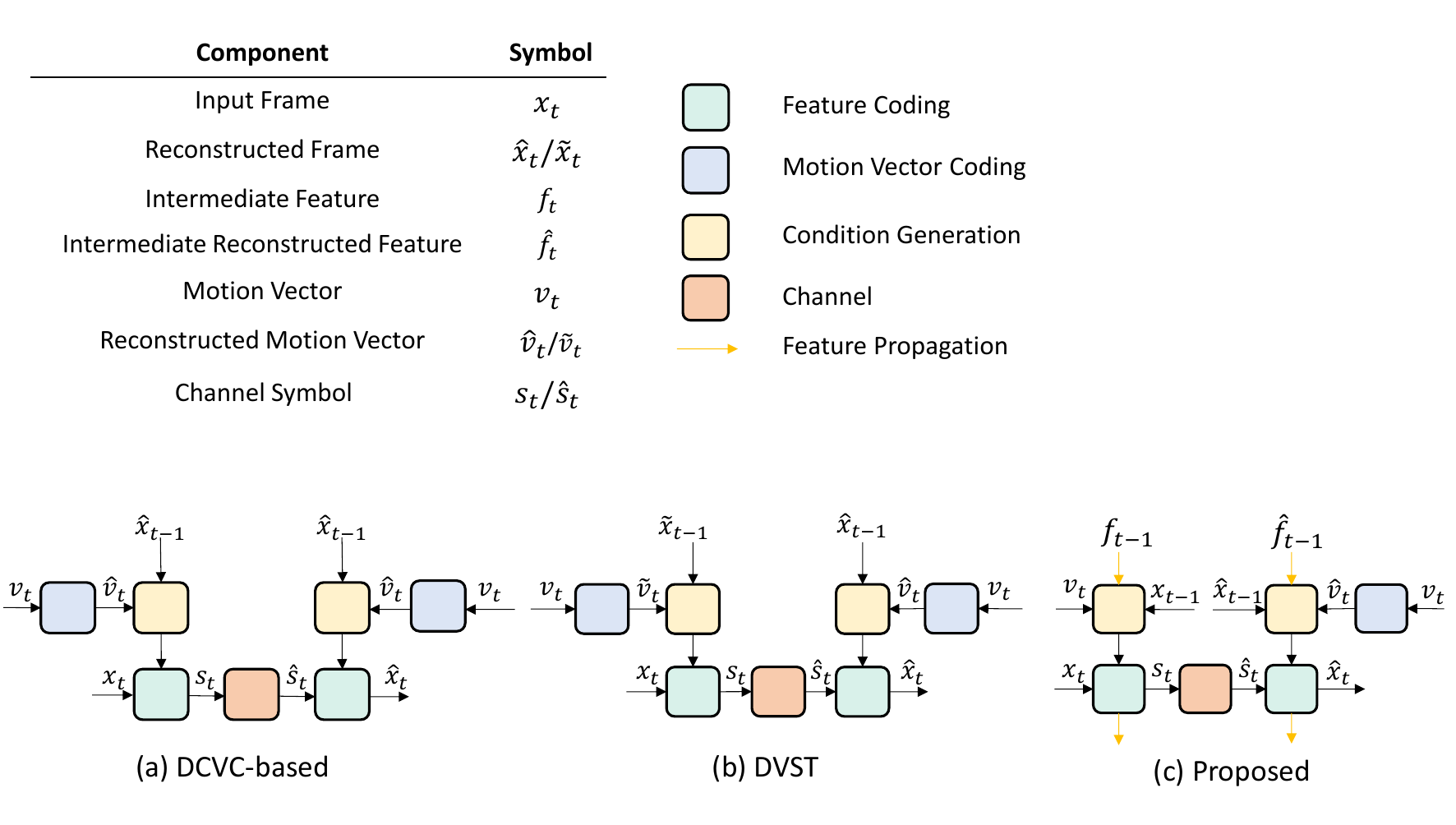}
\caption{Comparison for the relevant methods.}
\label{fig: comparasion}
\end{figure*}

\section{Proposed Method}
For a better understanding of the proposed scheme, a detailed comparison of several schemes is given in Fig. \ref{fig: comparasion}. Note that DVST \cite{dvst} and our method fall into DeepJSCC area while DCVC \cite{dcvc} belongs to deep source coding method. DCVC \cite{dcvc} itself is a digital compression scheme which uses the optical flow estimation network \cite{optic} to learn the motion vectors $v_t$ between the reference frame $x_{t-1}$ and the current frame $x_t$. After encoding and decoding the motion vectors, the reconstructed motion vectors $\hat{v}_t$ instruct the network from reference frames $\hat{x}_{t-1}$ to extract the context through a warping operation. The context is used as a condition to help the codec operation of the current frame. In DCVC \cite{dcvc}, the reconstructed video frames $\hat{x}_t$ and motion vectors $\hat{v}_t$ are naturally available at the encoder. For fair comparison, we use DCVC-based to denote the separated source-channel coding method which uses DCVC \cite{dcvc} concatenated with traditional channel coding.

DVST \cite{dvst} exploits nonlinear transform and conditional coding architecture in DCVC \cite{dcvc} to adaptively extract semantic features across video frames. The motion information $\tilde{v}_t/\hat{v}_t$ and the reference frame $\tilde{x}_{t-1}/\hat{x}_{t-1}$ generate the context by warping and convolution operations. The context guides the codec to further encode the features. Overall, it belongs to pseudo-analog transmission. The contexts can not be consistent even a pipeline of simulated transmission is constructed at the encoder which could also bring increase of time complexities and delay.
\begin{figure}[!t]
\center
\includegraphics[scale = 0.5]{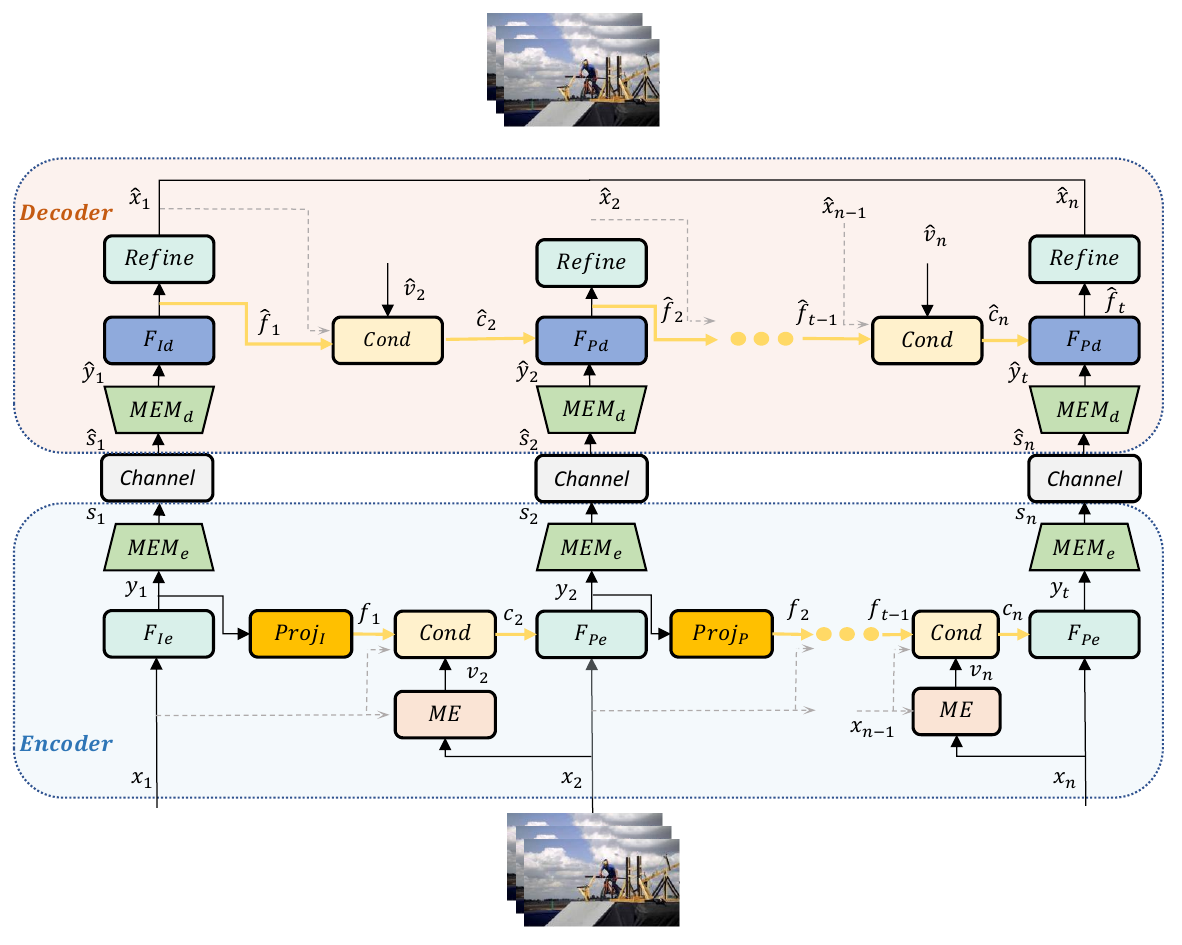}
\caption{Overall framework of the Proposed Schemes. Given an input frame $x_t$, the motion vector $v_t$ between $x_t$ and previous frame $x_{t-1}$ is estimated by $ME$. The condition generation network $Cond$ generates asymmetric contexts $c_t$ and $\hat{c}_t$ at the encoder and decoder, respectively. These contexts are subsequently fed into the feature encoder/decoder $F_{Pe}/F_{Pd}$ to perform the feature encoding and reconstruction of the current frame $x_t$. The gray dashed arrow indicates the propagation of the original and reconstruted frames. The black solid arrow represents the data flow at the encoder and decoder through the wireless channel. The yellow flow represents the propagation of the feature.}
\label{fig: framework}
\end{figure}

In our approach, we instruct the encoder to learn the coding conditions based on the ground-truth information $f_{t-1},x_{t-1}$ and ground-truth motion information $v_t$. That is, we do not need simulating the transmission pipeline at the encoder. We instruct the network learning coding from asymmetric context as the input of the conditional network at the encoder and decoder are ground-truth information $f_{t-1},x_{t-1},v_t$ and reconstructed information $\hat{f}_{t-1},\hat{x}_{t-1},\hat{v}_t$, respectively.

The overall problem statement is as follows. Considering a wireless video transmission problem with bandwidth and power constraints. The Group of Pictures (GOP) of video sequence $X = \{x_t\}_{t=1}^n$ includes $n$ frames, where each frame $x_t\in\mathbb{R}^{w\times h\times3}$ represents a vector of 24bit RGB pixel. Herein, $w$ and $h$ denote the width and height of the video frame, respectively. We design an encoding function $\Psi: \mathbb{R}^{n\times w\times h\times3}\rightarrow \mathbb{C}^{\sum{m_t}} $ to map the original video source $X$ to a set of complex channel symbol $S = \{s_t\}_{t=1}^n=\Psi(X)$, $s_t\in \mathbb{C}^{m_t}$, where $s_t$ represents $m_t$-dimensional encoding result of $x_t$.
After $S$ has been transmitted with the channel, the decoder takes the interfered symbols $\hat{S}$ as input and reconstructs the source $\hat{X}=\Upsilon(\hat{S})$ by the decoding function $\Upsilon: \mathbb{C}^{\sum{m_t}} \rightarrow \mathbb{R}^{n\times w\times h\times3}$.
Typically, $m_t<w \times h\times3$, $t\in\{1,...,n\}$, indicating bandwidth compression. We use the channel bandwidth ratio,
\begin{equation}
R=\frac{\sum_{t=1}^{n}m_t}{n\times w\times h\times3},
\end{equation}
to denote the average bandwidth required to transmit one pixel. 
In this paper, we mainly consider AWGN channel, where the transition function of the channel is defined as 
\begin{equation}
{\hat{S}}=S+N,
\end{equation}
and $N$ are samples following independent Gaussian distribution with zero mean $\mu$ and variance $\sigma^2$. The Channel Signal-to-Noise Ratio (CSNR) is defined as follows:\
\begin{equation}
CSNR = {10}\log_{10}{\frac{1}{\sigma^2}(dB)}.
\end{equation}

For other channel models, we only need to modify the channel transition function. In addition, we limit the power of the channel input to $P$:
\begin{equation}
\frac{1}{m_t}\mathbb{E}\left[\left|\left|s_t\right|\right|_2^2\right]\le P.
\end{equation}

\subsection{Overall Framework}
In general, we divide the video into GOP sequences, and one GOP includes $n$ consecutive frames of video. The first frame of the GOP represents the Intra-coded Frame (I-frame), which serves as a reference frame for coding the remaining Predictive-coded Frames (P-frames). The I-frame is coded using intra-frame coding, which is consistent with image coding. The P-frames, on the other hand, need to achieve higher compression ratios by capturing motion information and conditional coding. Accordingly, we need to design the encoding network and decoding network for I-frames and P-frames, respectively. The overall framework is shown in Fig. \ref{fig: framework}, where the first frame $x_1$ is an I-frame, and the corresponding transmission pipeline is the I-frame coding network. The remaining frames are P-frames, and their transmission pipelines correspond to the P-frame coding network. The details of them will be described later. The overall framework consists of four modules: feature encoder/decoder pairs $(F_{Ie/Pe}, F_{Id/Pd})$, motion estimation (ME) network, condition generation network $(Cond)$, and Mask-based Entropy Module (MEM), which includes an encoder $(MEM_e)$ and a decoder $(MEM_d)$.

\begin{figure*}[!t]
\center
\includegraphics[scale = 0.45]{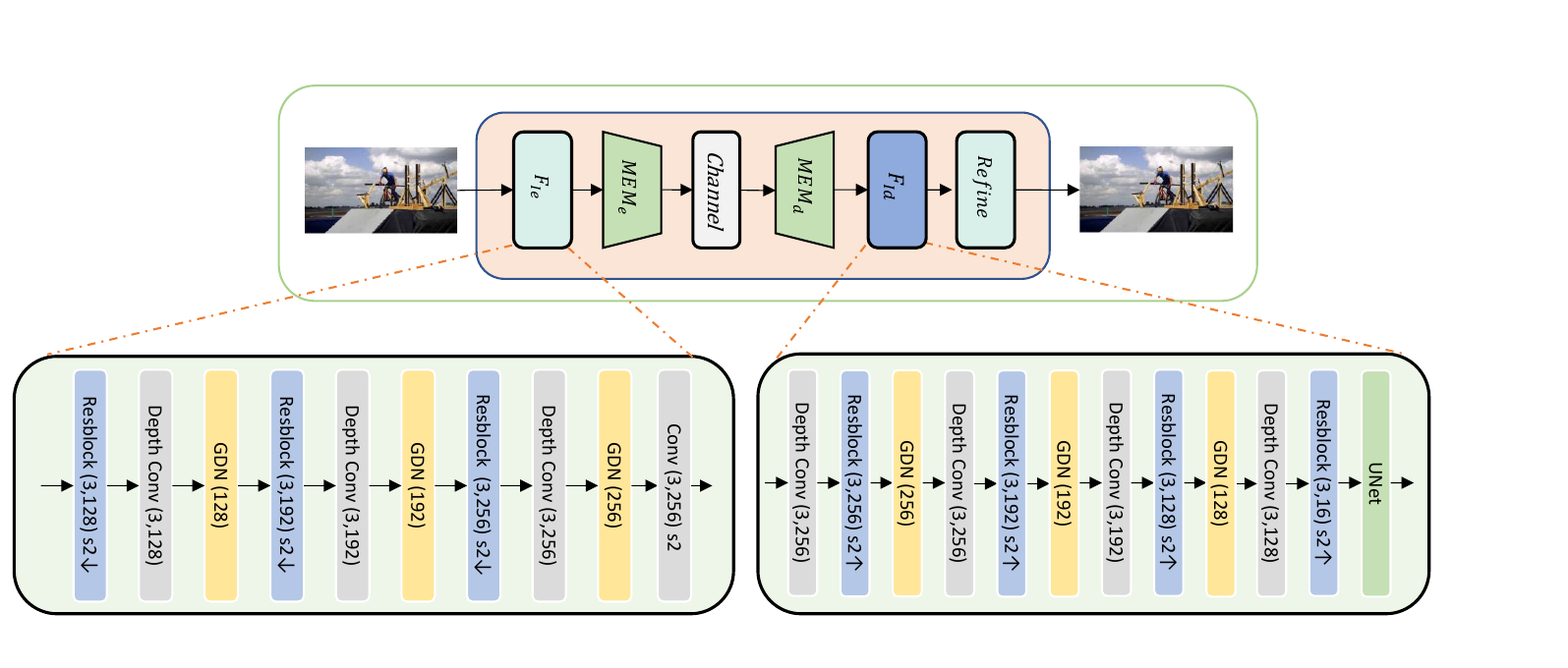}
\caption{Network architectures of I-frame encoder and decoder. The feature encoder $F_{Ie}$ extracts spatial features from the input frame, while the feature decoder $F_{Id}$ and $Refine$ reconstructs the frame at the decoder through feature decoding and refinement. $(M, N)$ represents the kernel size and number of output channels, $s2$ denotes stride of 2 and the followed $\uparrow/\downarrow$ indicates upsampling or downsampling. }
\label{fig: I-frame}
\end{figure*}

To encode a frame $x_t$, the image is first mapped to the latent space by $F_{Ie/Pe}$. Then, the variable-rate transmission is realized by $MEM_e$ based on the entropy of the latent variable. $s_t$, the result of $MEM_e$, is transmitted as the channel input. The channel symbols $s_t$ are defined as:

\begin{equation}
{s_t} = {MEM_{e}}({F_{Ie/Pe}}(x_t; c_t)),
\end{equation}
where $c_t$ denotes the encoding conditions during P-frame encoding process. After transmission through the wireless channel, the corresponding decoder $MEM_d$ receives the channel output $\hat{s}_t$, which firstly undergoes the reconstruction of the latent variable $\hat{y}_t$ done by $MEM_d$. The specifics of $MEM_e$ and $MEM_d$ will be developed in \ref{subsection 3B}. The recovered intermediate feature $\hat{f}_t$, is first obtained by $F_{Id/Pd}$:

\begin{equation}
{\hat{f}_t} = {F_{Id/Pd}}({MEM_{d}}(\hat{s}_t); \hat{c}_t),
\end{equation}
where $\hat{c}_t$ denotes the decoding conditions during P-frame decoding process. The intermediate features $\hat{f}_t$ are also propagated to the next P-frame during the decoding process through generating the conditions together with the reconstructed frames and motion information. Finally, the reconstruction current frame $x_t$ is generated by a $Refine$ network with multiple convolutional layers for enhancing the reconstructed frames:

\begin{equation}
{\hat{x}_t} = {Refine}(\hat{f}_t).
\end{equation}

Note that we additionally set up $Proj_{I/P}$ at the encoder to obtain the intermediate features $f_t$ which are propagated to the encoding of the next P-frame through generating the conditions together with the motion information estimated by the $ME$. The structure of $Proj_I$ for I-frame is different from that of $Proj_P$ for P-frame: $Proj_I$ consists of only one convolutional layer, while $Proj_P$ includes convolutional layers, Depth-separable Convolutional (Depth Conv) layers, ResNet \cite{resnet} Blocks, Generalized Normalization (GDN) layers and two U-Net networks. This design distinction is due to the fact that the intermediate features generated by $Proj_I$ do not include information from previous frames, whereas P-frames require the processing of information transmitted across multiple preceding frames, necessitating a more complex design.

\begin{figure*}[!t]
\center
\includegraphics[scale = 0.45]{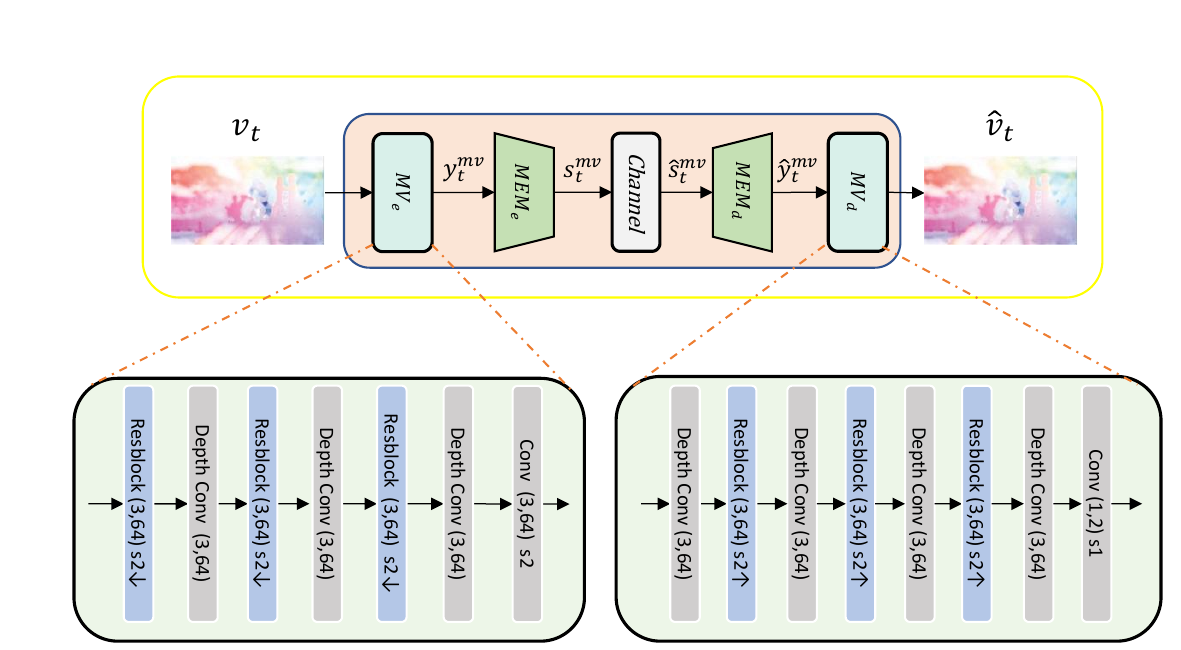}
\caption{Network architectures of the MV encoder and decoder. The MV encoder $MV_e$ and decoder $MV_d$ are responsible for encoding and decoding motion vectors respectively.}
\label{fig: MV-net}
\end{figure*}

\subsubsection{Coding Structure for I-frames}
The detailed encoding and decoding networks for I-frames are shown in Fig. \ref{fig: I-frame}. The networks are composed of ResNet blocks, Depth Conv layers and GDN layers. $F_{Id}$ uses the U-Net network as the last layer for reconstructed features enhancement. In order to increase the depth of the network and the receptive field of the convolution operation, we use the downsampled ResNet block and upsampled ResNet block in $F_{Ie}$ and $F_{Id}$, respectively. Also, we adopted the GDN layer, whose efficiency in the field of image compression and image density modeling has been demonstrated by a large number of research works \cite{GDN1, GDN2}. The GDN layer can significantly improve the training speed of the model. To further reduce the computational cost, we use the Depth Conv layer instead of the common-used 3D convolution. The use of Depth Conv layer allows for different numbers of channels for features with different resolutions. In addition, the entire Depth Conv layer structure consists of only convolution operations so that we can handle videos with different resolutions.

\begin{figure*}[!t]
\center
\includegraphics[scale = 0.45]{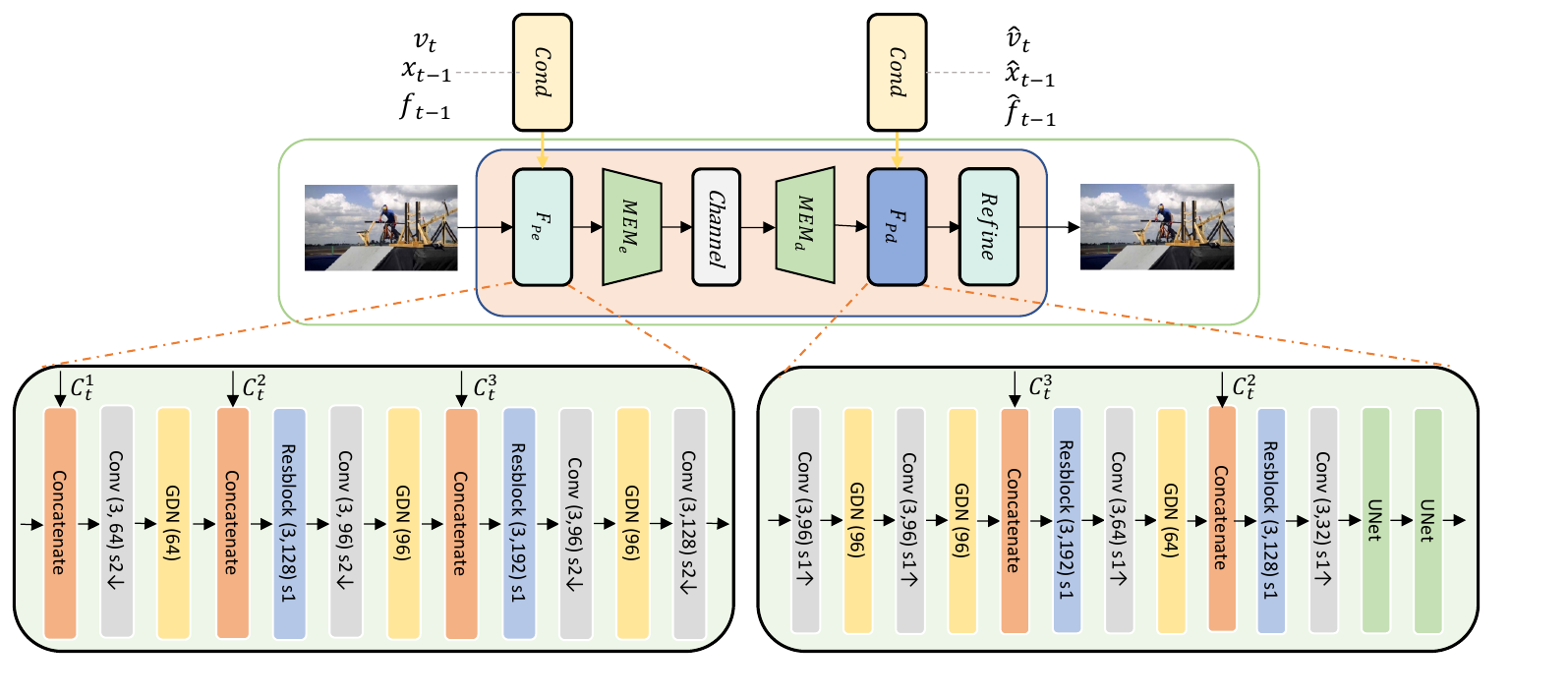}
\caption{Network architectures of the P-frame encoder and decoder. The asymmetric contexts are generated by the condition generation network $Cond$ at the encoder and decoder, and then fed into the corresponding feature encoder $F_{Pe}$ and decoder $F_{Pd}$.}
\label{fig: P-frame}
\end{figure*}

\subsubsection{Coding Structure for P-frames}
The encoder and decoder of P-frame are composed of convolutional layers, GDN layers, ResNet blocks. To enhance the quality of the decoded image, two additional U-Net networks are incorporated into $F_{Pd}$. In $F_{Pd}$, we employ sub-pixel convolution \cite{espcn} to implement convolution with upsampling capability. The detailed structure is shown in Fig. \ref{fig: P-frame}. To increase the coding efficiency for the $t$-th P-frame $x_t$ in a GOP, the contexts we adopt include the ground-truth previous frame $x_{t-1}$, the motion vector $v_t$ and the intermediate features $f_{t-1}$. Specifically, we first extract the motion vectors $v_t$ using a ME network:

\begin{equation}
{v_t} = {ME}(x_{t-1}, x_{t}).
\end{equation}

The $ME$ uses an optical flow estimation network \cite{optic}. Then the motion vectors $v_t$ are used to align $f_{t-1}$ with $x_{t-1}$ and generate conditions as close as possible. The coding contexts $C_t^1,C_t^2,C_t^3$ (scaled from small to large) with different scales are generated by $Cond$:

\begin{equation}
{C_t^1,C_t^2,C_t^3} = {Cond}(f_{t-1}, x_{t-1}, v_t).
\end{equation}

The $Cond$ adopts the diverse context extraction network in \cite{dcvc-dc}. Different from previous deep video compression, such diverse context extraction network can achieve conditional coding with multiple reference modes weighted by offset diversity. Specifically, features $f_{t-1}$ and reference frames $x_{t-1}$ are first aligned according to $v_t$ to obtain an initial reference frame. Then, a prediction network is employed to predict the residuals of multiple sets of motion vectors, referred to as offsets. The predicted residuals are combined with the motion vectors to obtain multiple sets of motion vectors, which are respectively combined with the features $f_{t-1}$ to generate the final coding conditions. Subsequently, the encoding condition is input to $F_{Pe}$ along with the current frame to mine the reference information provided by the context:

\begin{equation}
y_t = F_{Pe}({x_t; C_t^1,C_t^2,C_t^3}).
\end{equation}

Similarly, at the decoder, the decoding conditions $\hat{C}_t^1,\hat{C}_t^2,\hat{C}_t^3$ are generated and input to $F_{Pd}$:

\begin{equation}
{\hat{C}_t^1,\hat{C}_t^2,\hat{C}_t^3} = {Cond}(\hat{f}_{t-1}, \hat{x}_{t-1}, \hat{v}_t),
\end{equation}
\begin{equation}
\hat{f}_t = F_{Pd}({\hat{y}_t; \hat{C}_t^1,\hat{C}_t^2,\hat{C}_t^3}).
\end{equation}

It is worth noting that the conditions at the encoder and decoder are asymmetric in our framework. Therefore, as shown in Fig. \ref{fig: MV-net}, we set up a MV codec structure for $v_t$ which is similar to the one for I-frames. MV codec architecture is mainly composed of motion vector encoder/decoder pairs $(MV_e,MV_d)$ and MEM encoder/decoder pairs $(MEM_e,MEM_d)$. In the MV codec architecture, the motion vector $v_t$ is first mapped to the latent space by $MV_e$. The latent variable ${y^{mv}_t}$ is the output of $MV_e$. Then, the variable rate transmission is realized by $MEM_e$ based on the entropy of the latent variable ${y^{mv}_t}$: 

\begin{equation}
{s^{mv}_t} = {MEM_e}({MV_e}(v_t)),
\end{equation}
where ${s^{mv}_t}$ denotes the channel input symbol. The channel output $\hat{s}^{mv}_t$ are reconstructed by $MEM_d$ as the latent variable ${\hat{y}^{mv}_t}$ after passing through the wireless channel. Finally, the latent variable ${\hat{y}^{mv}_t}$ is decoded by $MV_d$ to get the decoded motion vector $\hat{v}_t$.

\begin{equation}
{\hat{v}_t} = {MV_d}({MEM_d}({\hat{s}^{mv}_t})).
\end{equation}

\begin{figure*}[!t]
\center
\includegraphics[scale = 0.5]{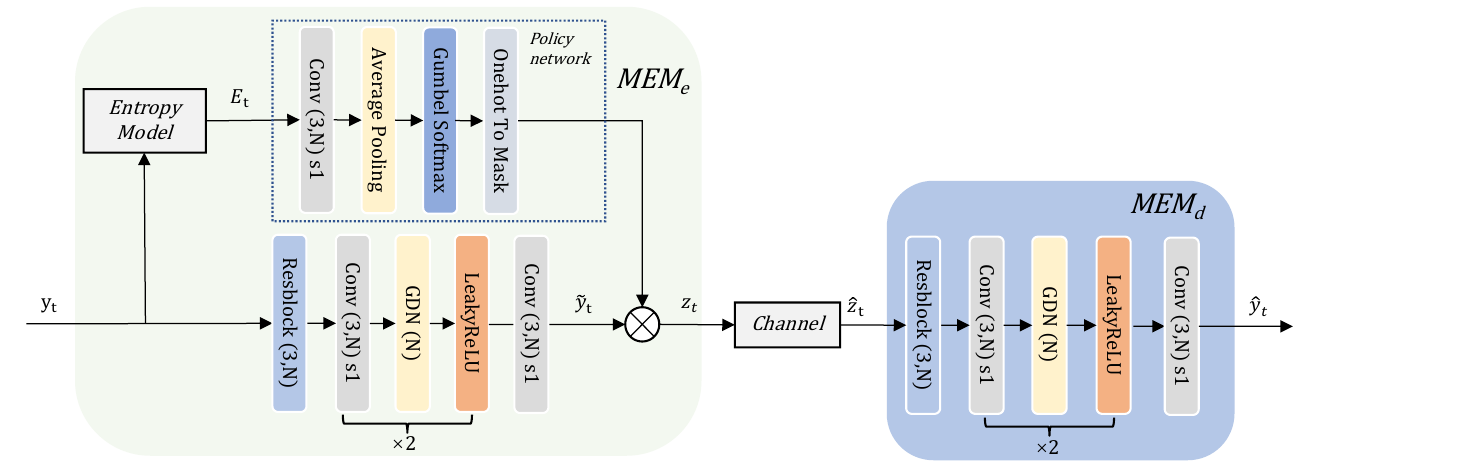}
\caption{Network architectures of the Mask-based Entropy Module. $E_t$ is the distribution parameter of the latent variable $y_t$ which is estimated by the entropy model. The policy network predicts the required bandwidth to transmit latent variables based on the estimated entropy and adjust the required bandwidth cost in the channel dimension by masking.}
\label{fig: mask-entropy}
\end{figure*}
 \begin{figure}[!t]
\center
\includegraphics[scale = 0.5]{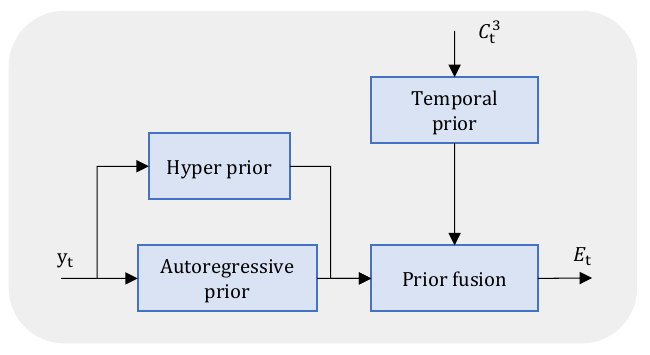}
\caption{Network architectures of the entropy model. The model fuses hyper, autoregressive and temporal priors to estimate the distribution parameters of latent variables.}
\label{fig: entropy model}
\end{figure}

\subsubsection{Feature Propogation}
As described in 3.1.2, similar to \cite{dcvc-dc}, the diverse contexts are utilized in our work for the condition generation; that is, we adopt intermediate features to provide reference information across multiple frames. In addition, we add the feature reconstruct term in our loss function which constrains the visual features of the previous reconstructed frame in the pixel domain and enables the network to capture useful information from different contexts. The previous reconstructed frame can only contain semantic information from the single frame, whereas the network intermediate features can provide more reference information due to fewer constraints, which can often get more information from multiple previous frames.

\subsection{Mask-based Entropy Module}\label{subsection 3B}
Considering that the feature maps output by the CNN have variability among different channels, we need to set up a variable-rate transmission mechanism for the latent variable $y_t$ for different channels. Flexibly adjusting the bandwidth requires transmitting $y_t$ without changing the structure of the network. In previous studies, it is common to fix the bandwidth required for transmission and then set the corresponding bandwidth allocation policy. This ignores the variability of video content. In traditional video compression, compression is achieved by controlling the entropy value, which also provides more dynamics in bandwidth utilization. Whereas in the JSCC system for video, the fixed bandwidth greatly limits the performance of the system. 

The variable-rate transmission mechanism in DVST \cite{dvst} ignores the information interaction among different channels of the latent variable, which will lose part of the encoded information and negatively impact the performance of video reconstruction. Considering this phenomenon, we propose the MEM, which adopts a dynamic strategy to generate masks based on the entropy of latent variables. Specifically, we first use the pre-trained entropy model to estimate the distribution parameters of latent variables and then utilize the policy network to predict the required bandwidth for transmitting latent variables based on the estimated entropy. Then, we adjust the required bandwidth cost in the channel dimension through the mask. Channels with low information entropy don't participate in the transmission and are represented as ``0" in the mask. This enables the encoder to dynamically adjust the bandwidth according to the video content to save transmission bandwidth while maintaining performance. In this way, the information of different channels of latent variables can interact with each other through the policy network, which reduces the loss of information during variable-length coding. Meanwhile, we only require a single network to process all channels, which significantly reduces memory and computational overhead compared to using multiple dynamic neural networks in DVST \cite{dvst}.

The network structure of the MEM is shown in Fig. \ref{fig: mask-entropy}. $y_t$ is the output of encoder $F_{Ie/Pe}$, and the mean $\mu_t$ and variance $\sigma^2_t$ of $y_t$ are estimated by the entropy model and concatenated to obtain $E_t=(\mu_t, \sigma^2_t)$. Then, $E_t$ is fed into a policy network to generate a mask $m_{t}$.
Each mask value $m_{t,i}$ corresponds to $i$-th channel of $y_t$, with $m_{t,i} \in \{0,1\}$. During training, we use Gumbel-Softmax \cite{Gumbel-Softmax} to generate one-hot vectors, making the training differentiable. In another branch, the latent representation $y_t$ is firstly processed through a series of convolutional layers to obtain $\tilde{y}_t$, which is then multiplied by the mask to produce the channel symbols $z_t$ for wireless transmission:
\begin{equation}
z_{t,i} = {\tilde{y}_{t,i}} \cdot {m_{t,i}},
\end{equation}
channels with $m_{t,i}=0$ are discarded and not transmitted through the wireless channel, effectively pruning low-entropy channels and reducing transmission bandwidth for content-adaptive transmission. $z_t$ is transmitted without a fraction of zeros, so we need to use channel coding and digital modulation to transmit the mask additionally to ensure lossless transmission and the bandwidth for transmitting the mask is much lower than that required for the video.

The effectiveness of MEM relies on the accurate entropy estimation provided by the entropy model. The entropy model serves as a density estimation model which can estimate the distribution parameters of latent variables including mean and variance \cite{balle2018}. It has shown excellent results in both conventional and deep video compression. We estimate distribution parameters of latent variables without requiring exact probabilities at the encoder or decoder using the entropy model, which eliminates the transmission of dictionaries in conventional coding. We concatenate the mean $\mu_t$ and variance $\sigma^2_t$ of the output of the entropy model to get $E_t=(\mu_t, \sigma^2_t)$. The architecture of the entropy model is shown in Fig. \ref{fig: entropy model}. Specifically, we use the hyper-prior module \cite{balle2018} to better extract more potential spatial correlations from the hyper-prior information. This information is then fused with the time-domain information extracted from the coding contexts $C_t^{3}$ with biggest scale and the distributional information extracted from the auto-regressive model to obtain the final distribution parameters. Additionally, there is only one I-frame and multiple consecutive P-frames in one GOP so that only the temporal prior information in P-frame network architecture is utilized in the prior fusion module.

\subsection{Loss function}
The training objective of the proposed system is to jointly optimize the bandwidth and distortion. Considering that within one GOP, I-frame and P-frames are coded independently; that is, P-frames coder utilizes the conditions while I-frame coder does not, we train the I-frame and P-frame coder separately. 
 
In the first stage, we train the I-frame coder. As there is no intermediate features participated into the coding of I-frame, we can follow the optimization objective used in the common DeepJSCC-based image transmission scheme \cite{djscc4,djscc5} to optimize the I-frame coding network with the following loss function:
\begin{equation}
L_1=\lambda R_t+D_t,
\end{equation}
where $R_t$ is the channel bandwidth ratio of the t-th I-frame which is computed by the mask generated by the policy network in MEM. $D_t$ denotes the Mean Squared Error (MSE) between the original frame and the reconstructed frame. $\lambda$ is a hyperparameter to controls the trade-off between the bandwidth and the distortion. 

In the second stage, we fix the I-frame coder and train the P-frame coder and motion vector coder. Inspired by \cite{DCVC-HEM, DCVC-TCM, m-lvc}, we adopt different weighting parameters for each P-frame encoder to help them converge faster to achieve better performance. Details can be found in Section \ref{sec:experiment}. We update the network frame by frame using the following loss function:
\begin{equation}
L_2=\lambda R_t+ w_t(D_t+D_{f_t}) .
\end{equation}
The distortion term includes not only the reconstruction distortion of video frames $D_t$ but also the distortion of intermediate propagated features $D_{f_t}$, enabling the network to capture useful information from different contexts. Besides, we notice that due to error propagation during frame-by-frame video compression, the distortion of latter frames becomes more and more serious. So we introduce $w_t$ as a weighting factor for distortion, which mitigates error propagation by adjusting the distortion of each frame. 
As in the previous schemes \cite{dcvc, dcvc-dc, dvc3}, we update the model step by step to make the training more stable. Similar to commonly-used training strategies in recent papers \cite{DCVC-TCM, nvc-gan, nvc-cross}, in the last 5 epochs, we update the network parameters on the GOP sequence using the following loss function to further reduce error propagation.
\begin{equation}
L=\sum_{t=1}^{N}{(\lambda}R_t+w_t D_t) .
\end{equation}

\begin{figure*}[htbp]
\centering
\subfigure[HEVC Class B ($1920 \times 1080$)]{
\includegraphics[width=0.4\linewidth]{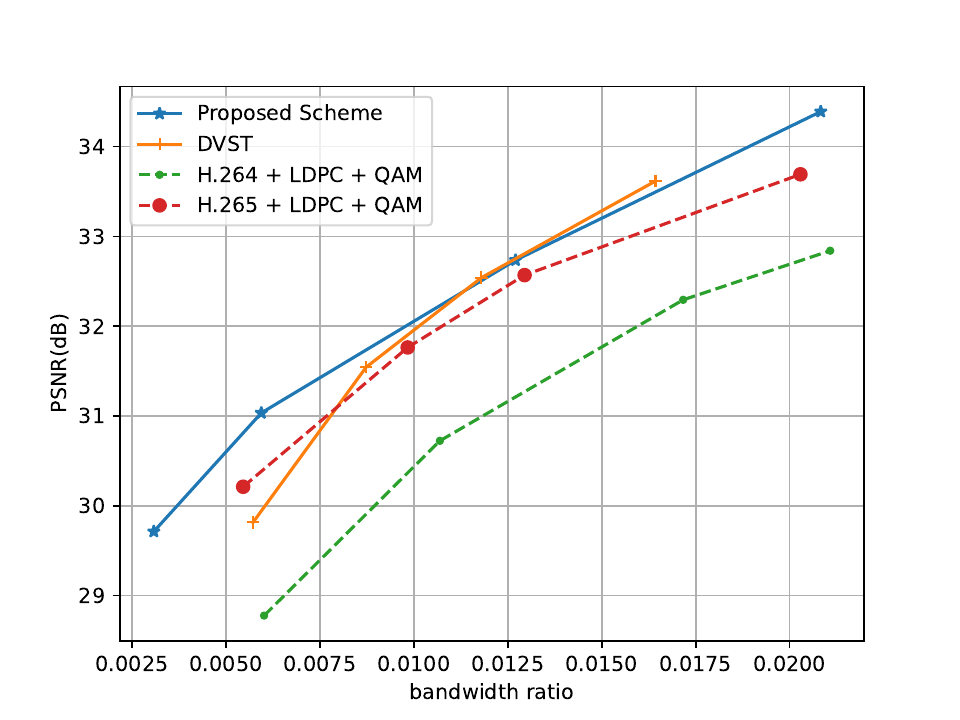}}
\hspace{-0.02\linewidth}
\subfigure[HEVC Class C ($832 \times 480$)]{
\includegraphics[width=0.4\linewidth]{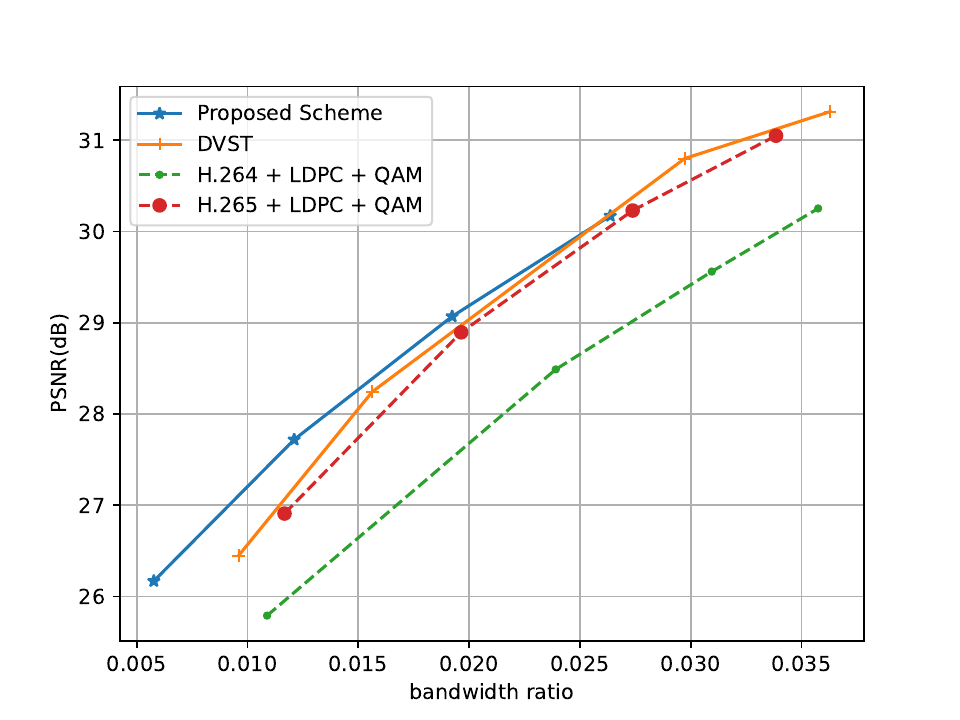}}
\hspace{-0.02\linewidth}
\subfigure[HEVC Class D ($416 \times 240$)]{
\includegraphics[width=0.4\linewidth]{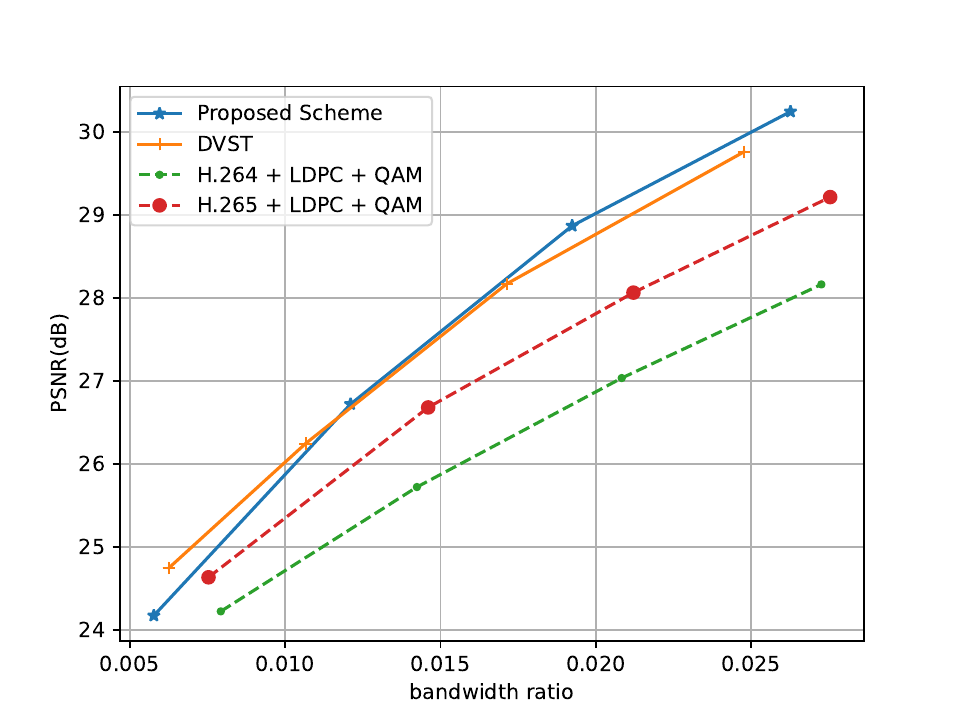}}
\hspace{-0.02\linewidth}
\subfigure[HEVC Class E ($1280 \times 720$)]{
\includegraphics[width=0.4\linewidth]{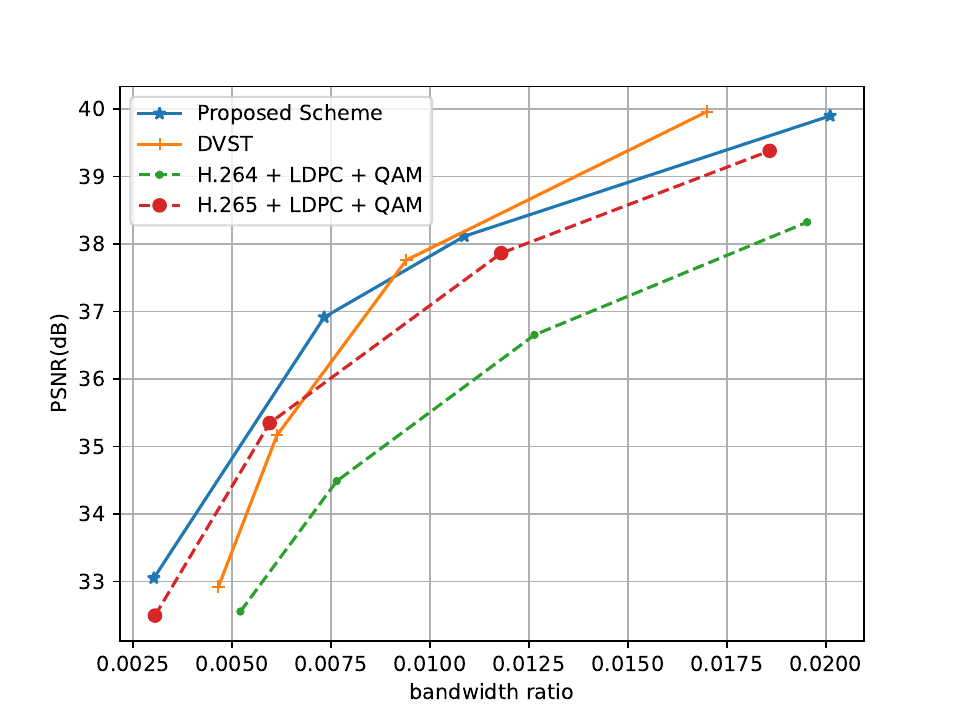}}
\caption{PSNR performance versus the average bandwidth ratio of the relevant schemes over the AWGN channel (CSNR = 10dB, GOP = 4).}
\label{fig: gop4psnr}
\end{figure*}

\begin{figure*}[!t]
\centering
\subfigure[HEVC Class B ($1920 \times 1080$)]{
\includegraphics[width=0.4\linewidth]{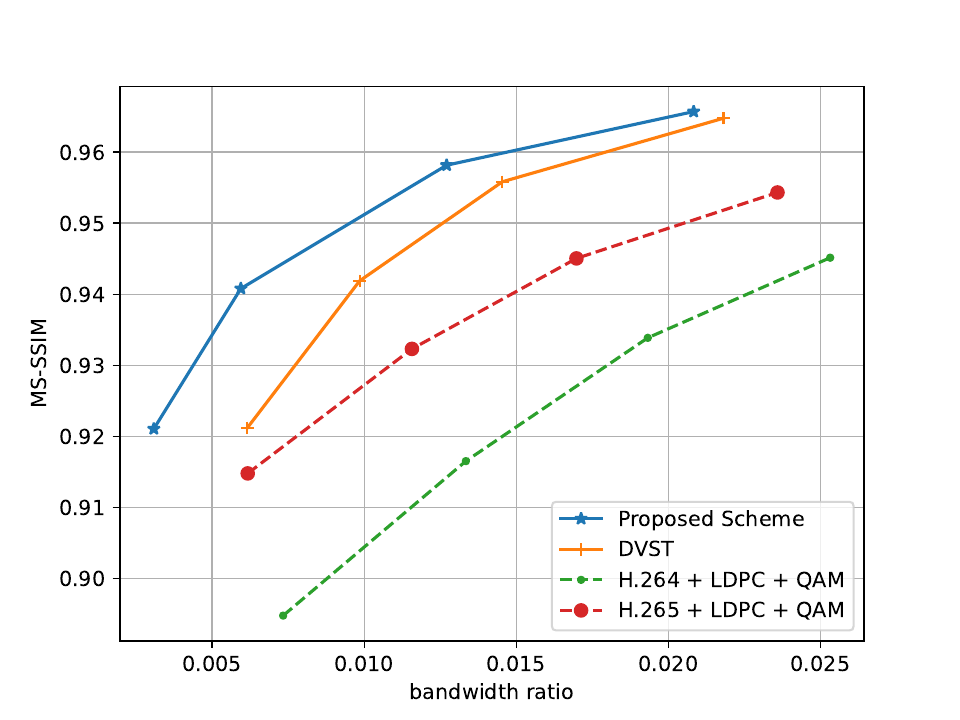}}
\hspace{-0.02\linewidth}
\subfigure[HEVC Class C ($832 \times 480$)]{
\includegraphics[width=0.4\linewidth]{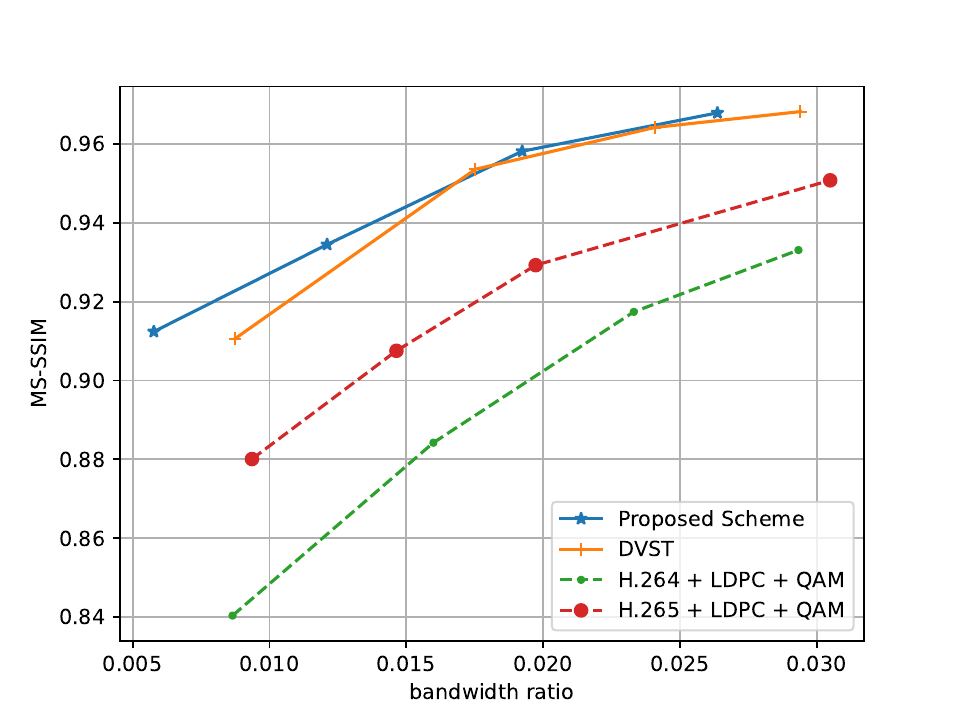}}
\hspace{-0.02\linewidth}
\subfigure[HEVC Class D ($416 \times 240$)]{
\includegraphics[width=0.4\linewidth]{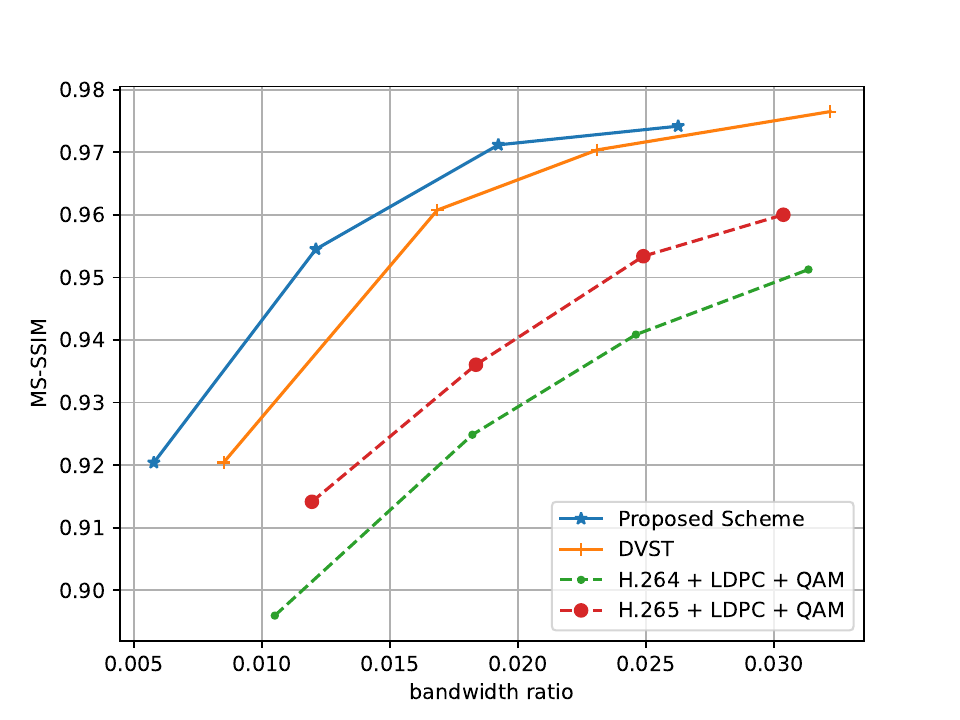}}
\hspace{-0.02\linewidth}
\subfigure[HEVC Class E ($1280 \times 720$)]{
\includegraphics[width=0.4\linewidth]{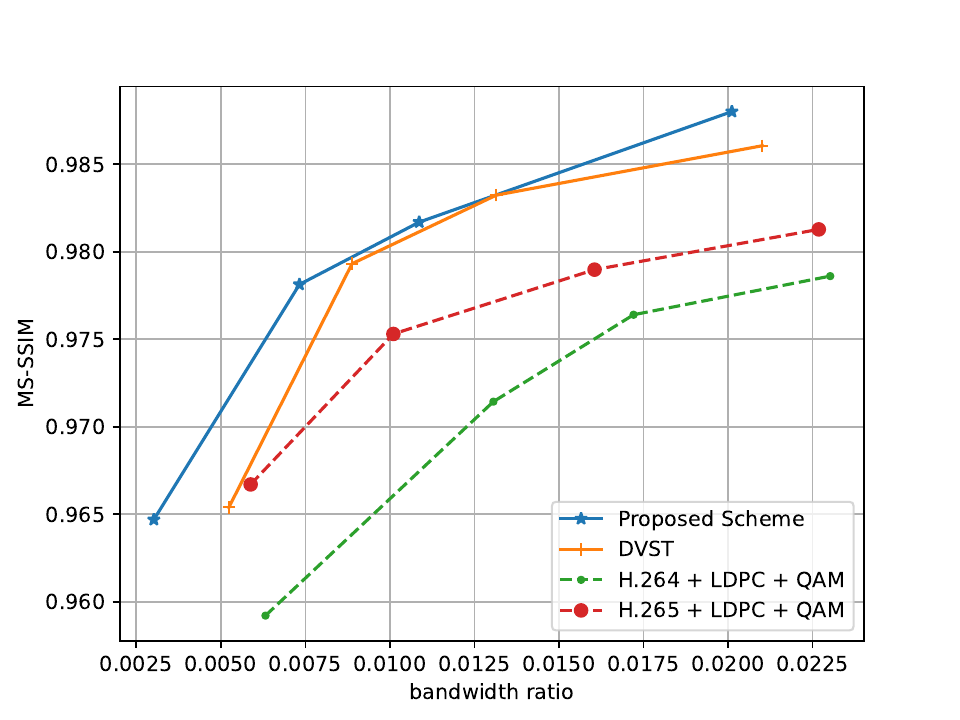}}
\caption{MS-SSIM performance versus the average bandwidth ratio of the relevant schemes over the AWGN channel (CSNR = 10dB, GOP = 4).}
\label{fig: gop4ssim}
\end{figure*}

\section{Experiments}
\label{sec:experiment}
\subsection{Setup}
\begin{itemize}
\item{Datasets:} We first randomly selected 500,000 images from the Imagenet dataset \cite{imagenet} for training the I-frame coder. For training the P-frame coder, we used the widely adopted Vimon-90k dataset \cite{vimon} as in most research, and divided it into training and testing sets. During training, the images were randomly cropped to a size of $256\times256$. For the testing, we adopt the HEVC \cite{hevc} dataset and the MCL-JCV \cite{mcl} dataset to evaluate the performance and generalization of our model. These datasets are also commonly used in the field of video coding, containing diverse visual quality content, such as targets with high-speed and low-speed motion. The HEVC dataset includes 16 YUV420-format videos of four types. These videos not only have different styles but also varying resolutions. The MCL-JCV dataset comprises 30 1080P videos.
\item{Metrics:} To measure the channel variations, we utilize the bandwidth compression ratio and CSNR. As for the video reconstruction quality, we employ the Peak Signal-to-Noise Ratio (PSNR) and Multi-Scale Structural Similarity Index (MS-SSIM). In addition, to better reflect perceptual quality consistent with human visual perception, we further adopt the Learned Perceptual Image Patch Similarity (LPIPS) \cite{lpips} metric as a complementary evaluation criterion.
\end{itemize}

To measure the transmission efficiency, we define the channel bandwidth ratio $R_t$ to quantify the proportion of symbols transmitted over the wireless channel relative to the source data size. In the proposed MEM, latent channels with low information entropy do not participate in the transmission and are represented as zeros in the mask. For a frame with spatial resolution $h \times w$, each latent feature map corresponding to the $i$-th channel has a spatial size of $h/16 \times w/16$, which results from the cumulative downsampling effect of four convolutional layers with a stride of 2 in the P-frame network, as illustrated in Fig. \ref{fig: P-frame} and Fig. \ref{fig: mask-entropy}. Therefore, the actual channel bandwidth ratio of the $i$-th frame is depicted as:
\begin{equation}
R_t=\frac{\sum_{t=1}^{C}{m_{t,i} \times {h/16} \times {w/16}}}{3 \times h \times w}, m_{t,i} \in \{0,1\},
\end{equation}
where $C$ denotes the total number of latent channels. The overall average bandwidth ratio for one GOP is defined as:
\begin{equation}
R={\frac{1}{n}}{\sum_{t=1}^{n}R_t}.
\end{equation}

We first trained four models with different values of $\lambda=(4e-3, 3e-3, 2e-3, 1e-3)$ under CSNR = 10dB, corresponding to different bandwidth ratios. It is noteworthy that the same $\lambda$ is used during the three stages of model training. We employed the Adam optimizer with a learning rate of $1e-4$ to update the model parameters during training. For the I-frame encoder, we trained with a batch size of 32 on one RTX 4090 GPU. For the P-frame encoder, we used a batch size of 2. We set the loss weights for P-frames according to the literature \cite{dcvc-dc}. Each sequence in the Vimon-90k dataset contains only 7 frames, and we simultaneously trained four P-frames for each sequence. Therefore, the weights $w_t$ for the four P-frames are set to $(0.5, 1.2, 0.9, 1.2)$.

\begin{figure*}[htbp]
\centering
\subfigure[HEVC Class B ($1920 \times 1080$)]{
\includegraphics[width=0.4\linewidth]{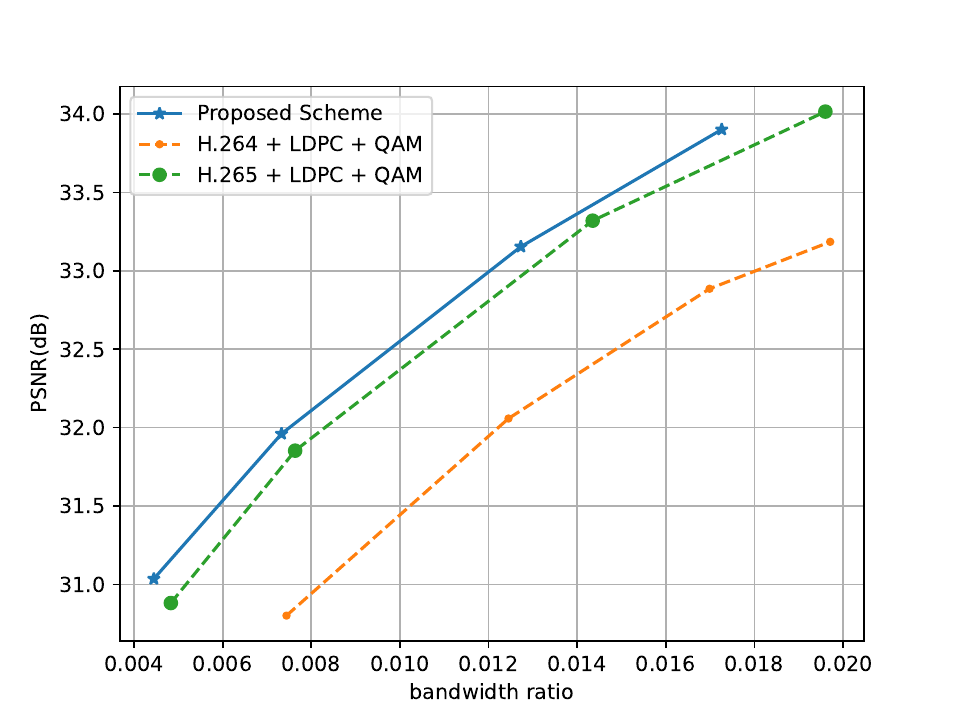}}
\hspace{-0.02\linewidth}
\subfigure[HEVC Class C ($832 \times 480$)]{
\includegraphics[width=0.4\linewidth]{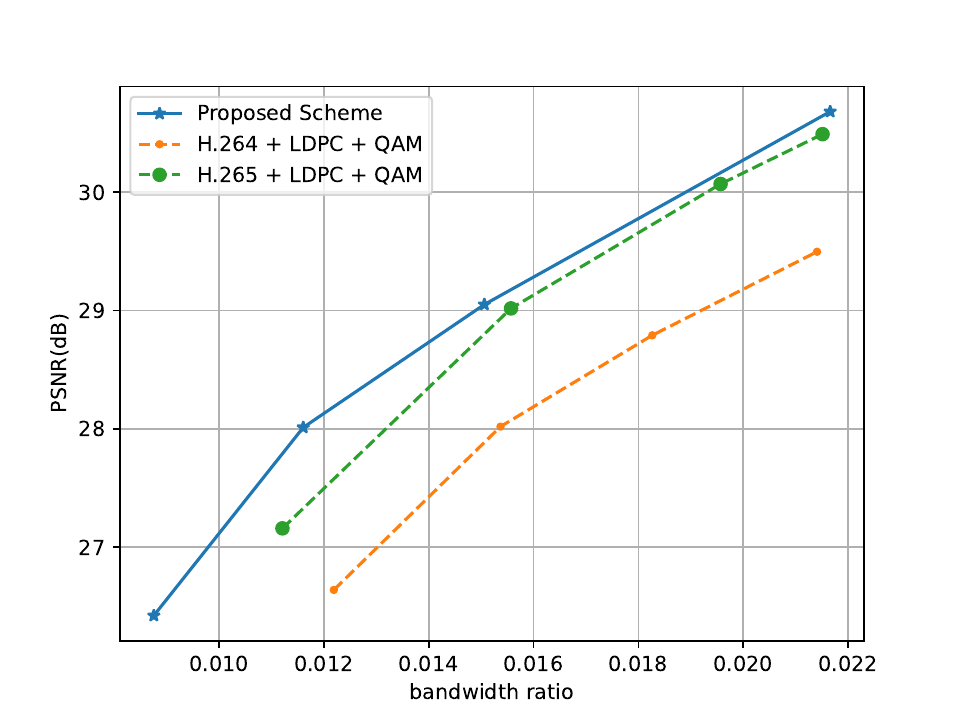}}
\hspace{-0.02\linewidth}
\subfigure[HEVC Class D ($416 \times 240$)]{
\includegraphics[width=0.4\linewidth]{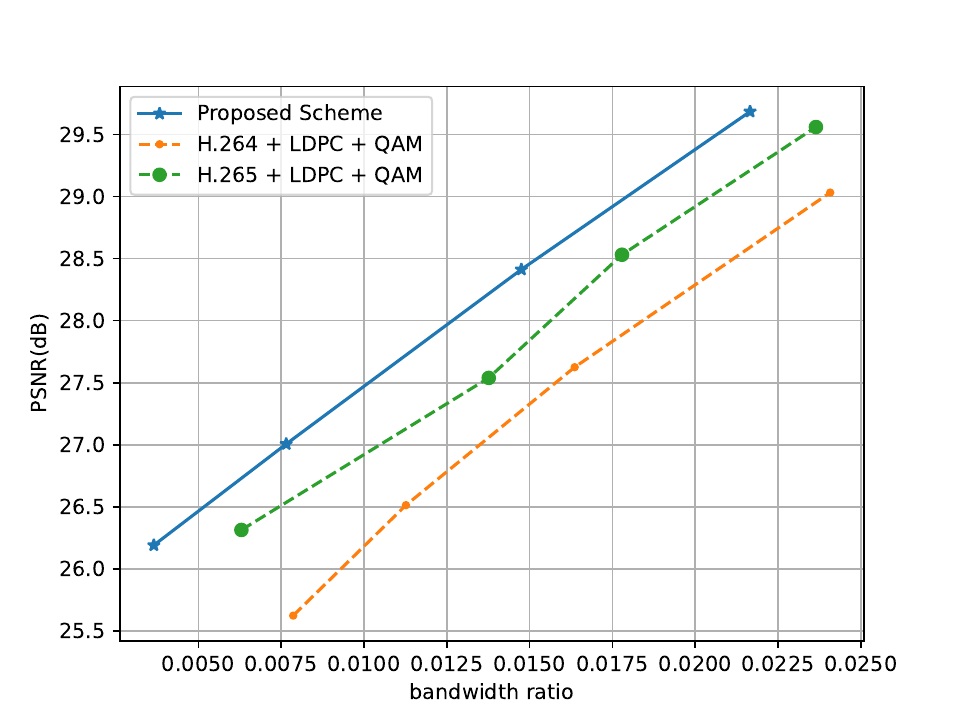}}
\hspace{-0.02\linewidth}
\subfigure[MCL-JCV ($1920 \times 1080$)]{
\includegraphics[width=0.4\linewidth]{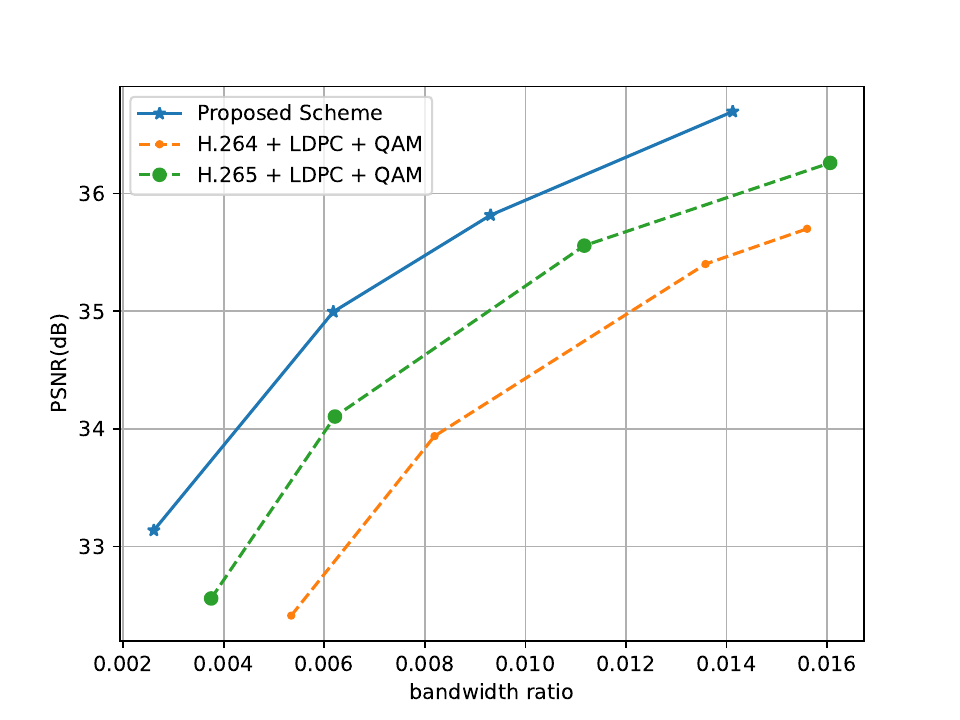}}
\caption{PSNR performance versus the average bandwidth ratio of the relevant schemes over the AWGN channel (CSNR = 10dB), GOP = 10 for HECV sequence and GOP = 12 for MCL-JCV sequence.}
\label{fig: gop12psnr}
\end{figure*}

\begin{figure*}[!t]
\centering
\subfigure[HEVC Class B ($1920 \times 1080$)]{
\includegraphics[width=0.4\linewidth]{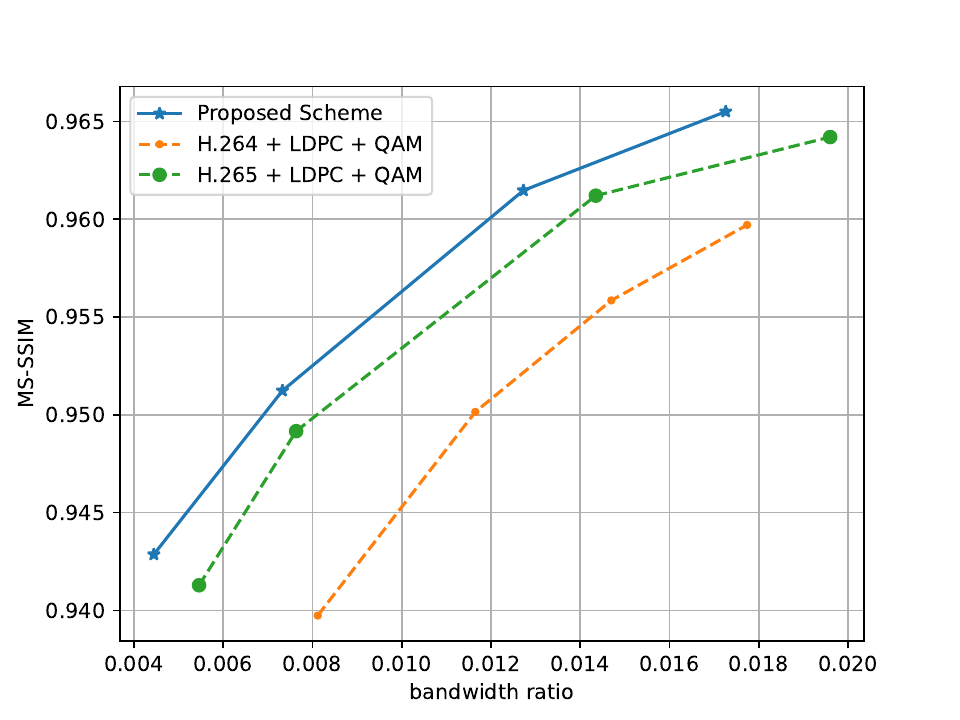}}
\hspace{-0.02\linewidth}
\subfigure[HEVC Class C ($832 \times 480$)]{
\includegraphics[width=0.4\linewidth]{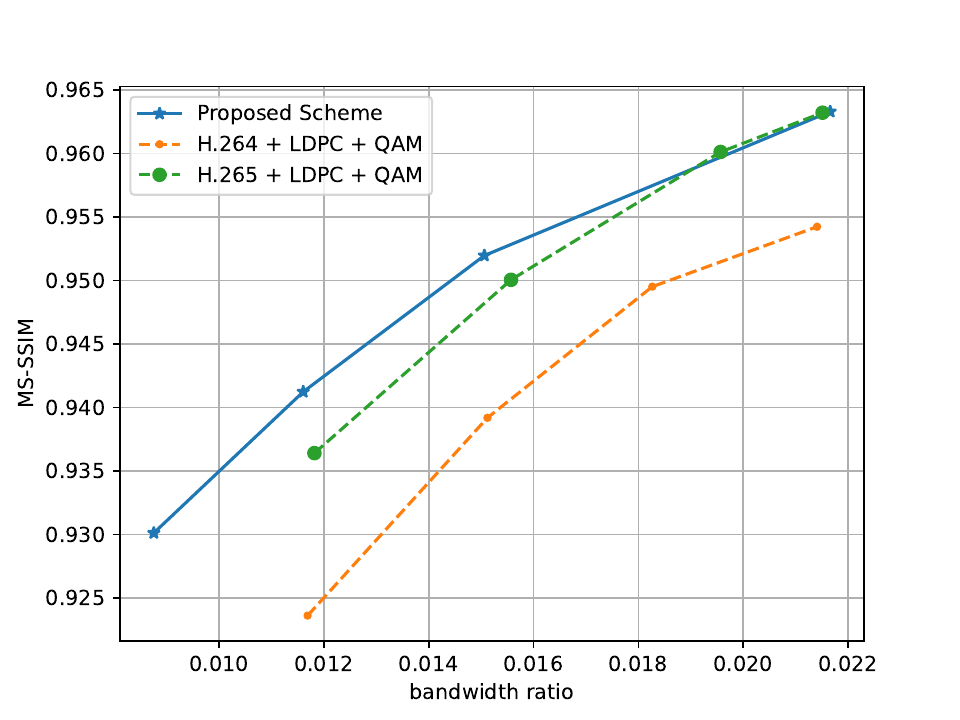}}
\hspace{-0.02\linewidth}
\subfigure[HEVC Class D ($416 \times 240$)]{
\includegraphics[width=0.4\linewidth]{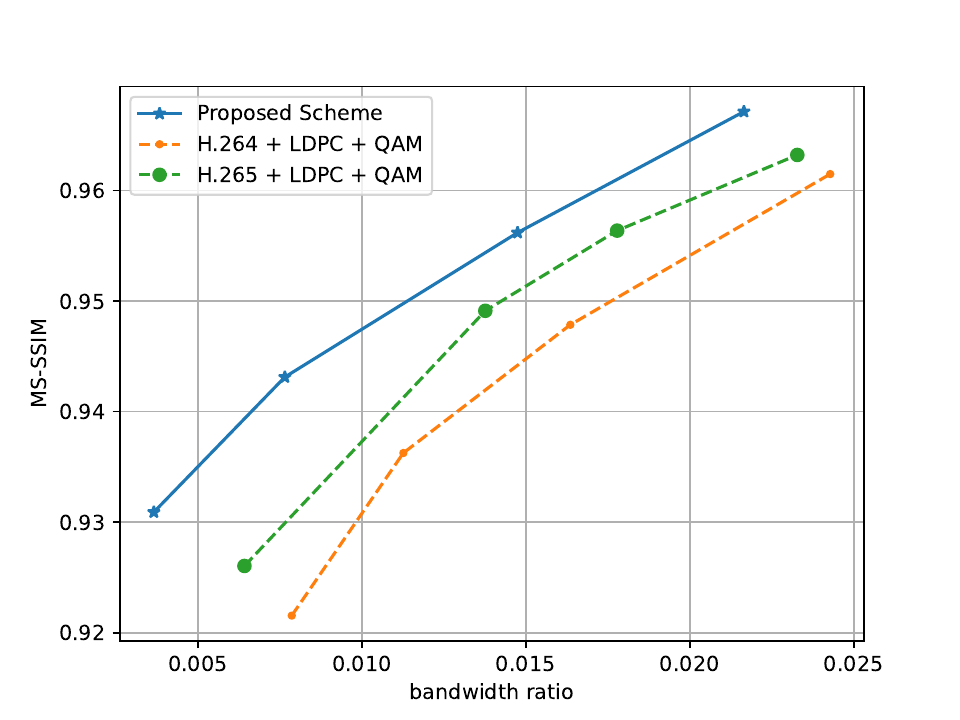}}
\hspace{-0.02\linewidth}
\subfigure[MCL-JCV ($1920 \times 1080$)]{
\includegraphics[width=0.4\linewidth]{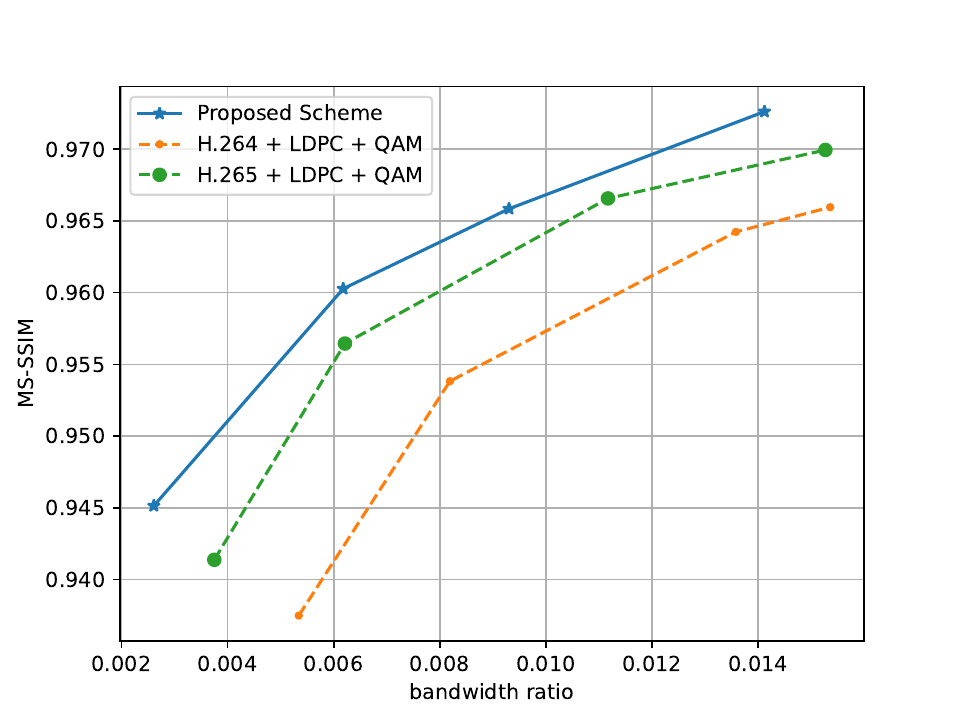}}
\caption{MS-SSIM performance versus the average bandwidth ratio of the relevant schemes over the AWGN channel (CSNR = 10dB), GOP = 10 for HECV sequence and GOP = 12 for MCL-JCV sequence.}
\label{fig: gop12ssim}
\end{figure*}

\begin{figure}[!t]
\centering
\subfigure[HEVC Class B ($1920 \times 1080$)]{
\includegraphics[width=0.45\linewidth]{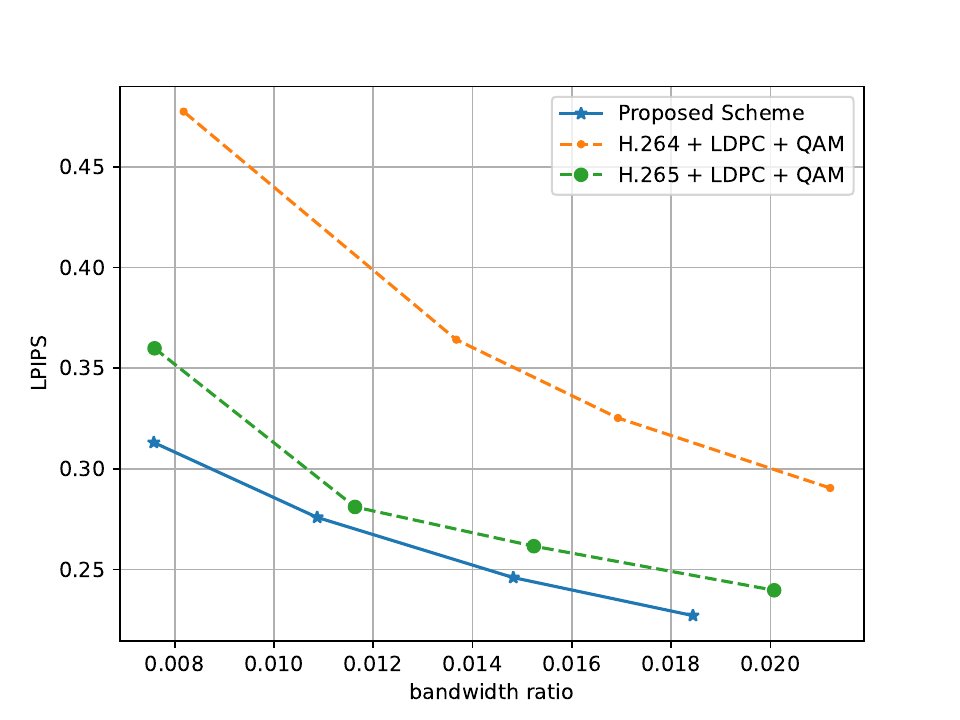}}
\subfigure[HEVC Class E ($1280 \times 720$)]{
\includegraphics[width=0.45\linewidth]{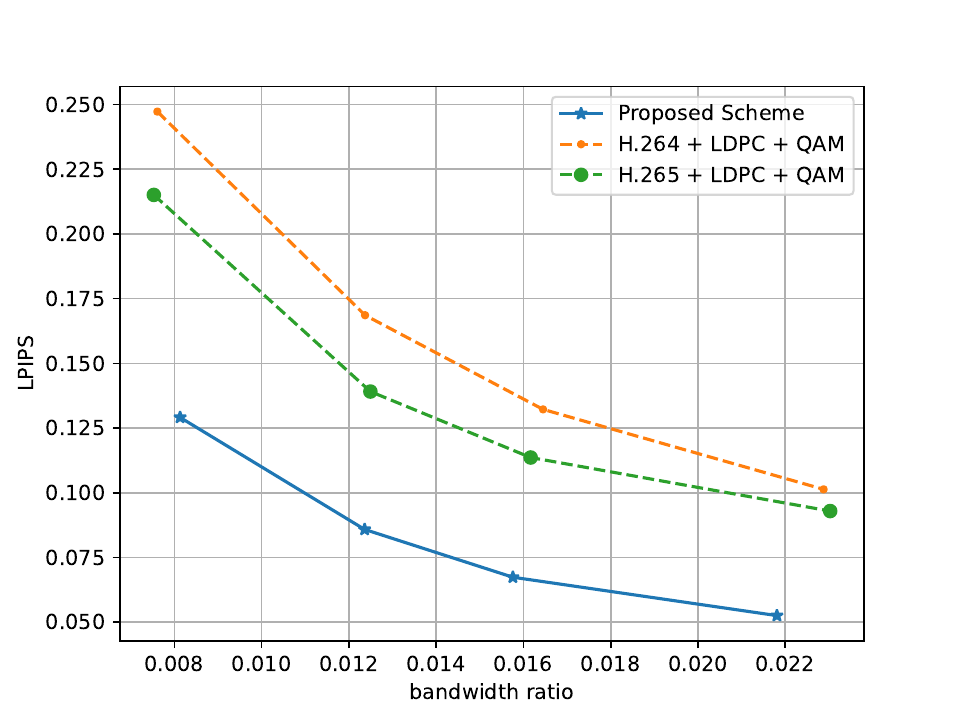}}
\caption{LPIPS performance versus the bandwidth ratio of relevant schemes over the AWGN channel (CSNR = 10dB), GOP = 10 for HEVC Class B and Class E test sequence.}
\label{fig: gop12lpips}
\end{figure}

The classic video transmission systems are employed as baselines, which include H.264-based, and H.265-based video transmission schemes. Additionally, we also adopted DVST \cite{dvst} as the baseline. For the channel encoding and modulation strategies in the separate framework, we utilized LDPC and QAM, respectively. As for channel model, we consider the widely used AWGN channel with channel SNR = 10dB as same as DVST \cite{dvst}. By combining these source coding, channel coding, and modulation schemes, we constructed the digital communication scenarios at various rates. For H.264 and H.265, we followed the settings in \cite{dvst}, configuring FFmpeg to the ``veryfast" and low-latency mode. The difference lies in using RGB as the input space.

\subsection{Results}
First, we set the GOP to 4, and compared our approach with the latest deep JSCC-based video transmission scheme, DVST \cite{dvst}. Fig. \ref{fig: gop4psnr} illustrates the rate-distortion (R-D) performance curves of various schemes with PSNR as the metric under CSNR 10dB. Herein, ``H.26x+LDPC+QAM" represents the combination of H.26x, LDPC, and QAM as the entire transmission system compared with our approach. 

For H.26x + LDPC + QAM, after traversing given combinations of LDPC coding rates and QAM modulation schemes, we employed 2/3 rate LDPC (block length between 4096 and 6144) and 16QAM under 10dB CSNR to ensure reliable transmission for the digital signals. From Fig. \ref{fig: gop4psnr}, we observe that our scheme outperforms the digital-based methods for various types of videos in the HEVC standard dataset. Furthermore, under certain bandwidth constraints, our scheme exhibits superior performance compared to DVST \cite{dvst}. Our scheme can improve quality correspondingly with the increase in channel bandwidth. Additionally, our approach performs better in HEVC Class B and Class E compared to HEVC Class C and Class D. This is because our Content-adaptive variable bandwidth transmission implementation offers better performance gains in videos with higher resolutions. The video of Class B and Class E have relatively higher resolutions and smoother content, with less complex motion compared to Class C and Class D, making it challenging for our scheme to learn encoding based on different contexts.

It's worth noting that the bandwidth compression ratio of our proposed scheme may fluctuate within different ranges for models trained under different $\lambda$. Additionally, the model can adjust the transmission bandwidth accordingly for different types of videos. Fig. \ref{fig: gop4ssim} illustrates the performance of various approaches using MS-SSIM as the evaluation metric. MS-SSIM, compared to PSNR, better aligning with human visual perception, often exhibits superior performance in the field of computer vision. Herein, we did not train our model using MS-SSIM as the distortion metric, yet we achieve better results compared to DVST \cite{dvst}. In most test datasets, our approach demonstrates the best R-D performance.

\begin{figure*}[hbtp]

\centering
\subfigure[]{
\includegraphics[width=0.46\linewidth]{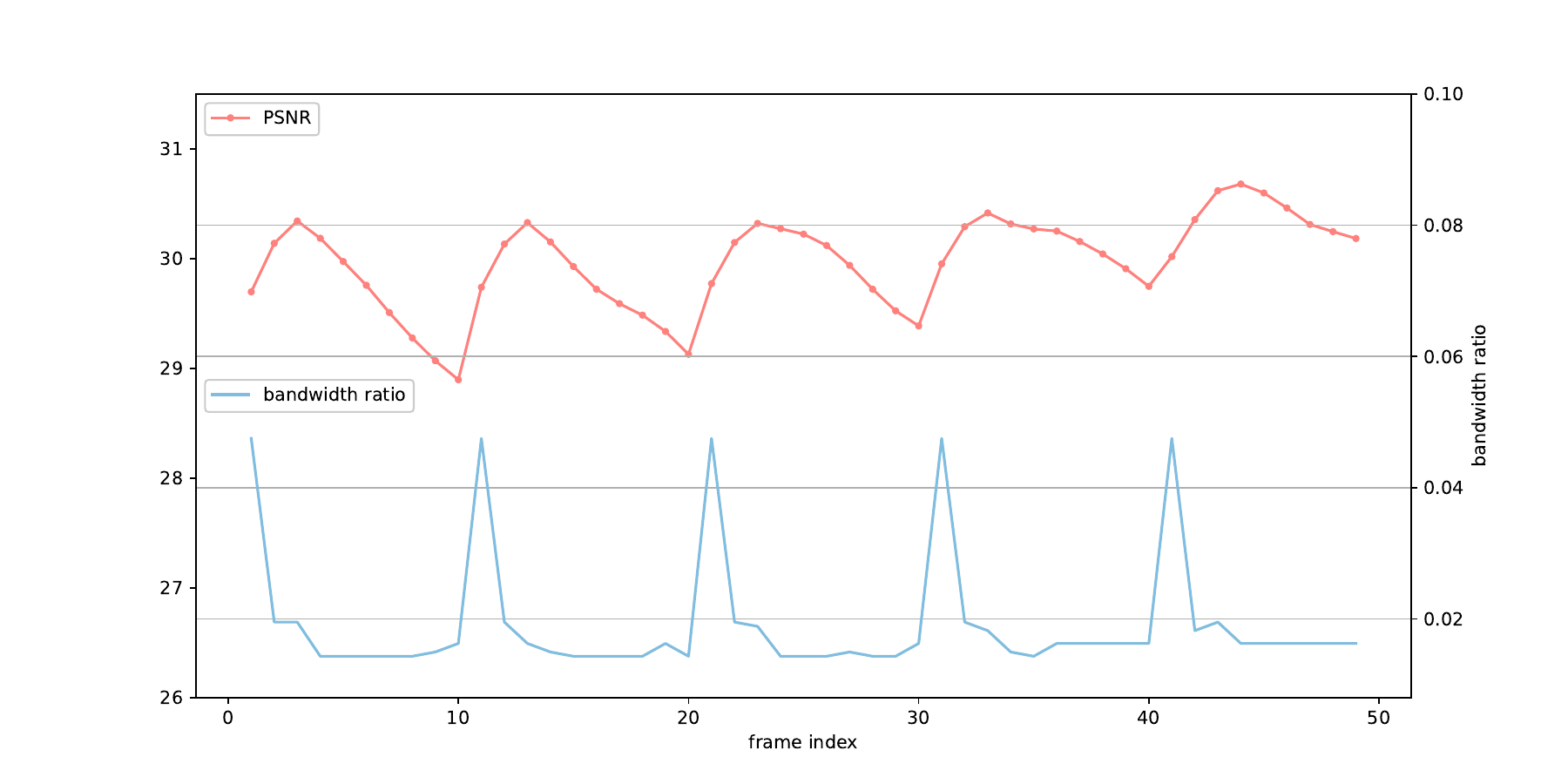}}
\hspace{-0.02\linewidth}
\subfigure[]{
\includegraphics[width=0.46\linewidth]{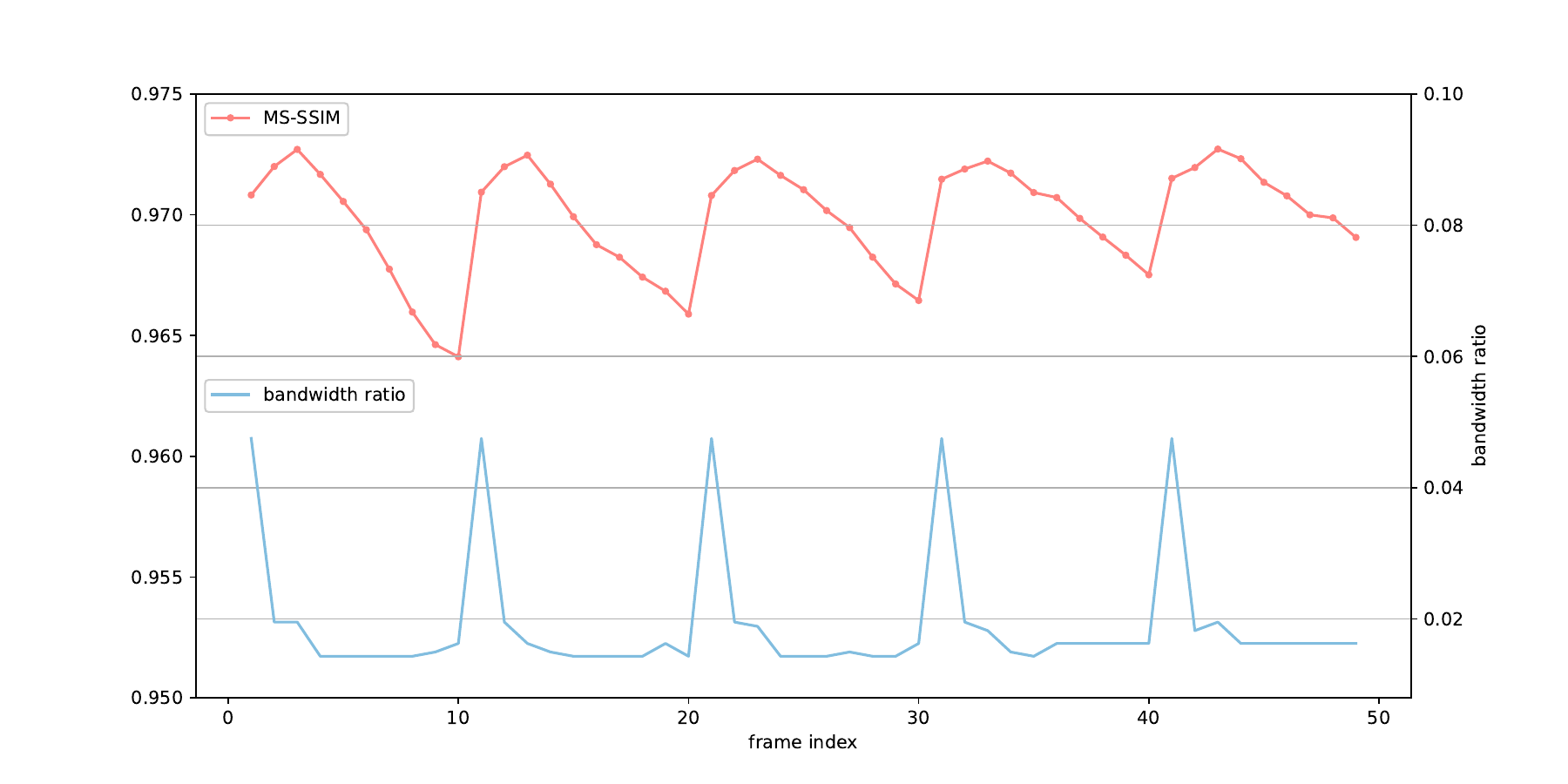}}
\subfigure[]{
\includegraphics[width=0.46\linewidth]{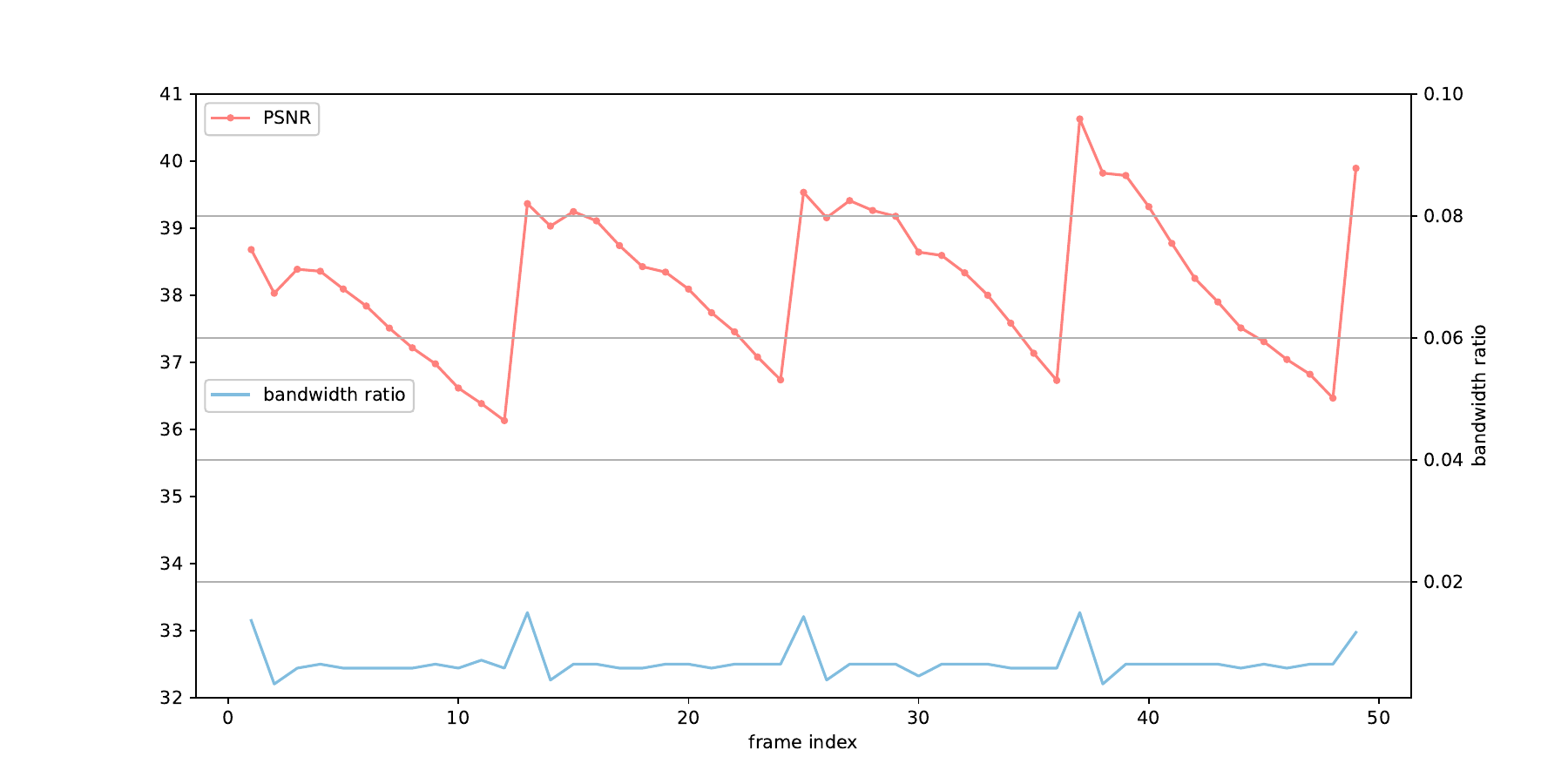}}
\hspace{-0.02\linewidth}
\subfigure[]{
\includegraphics[width=0.46\linewidth]{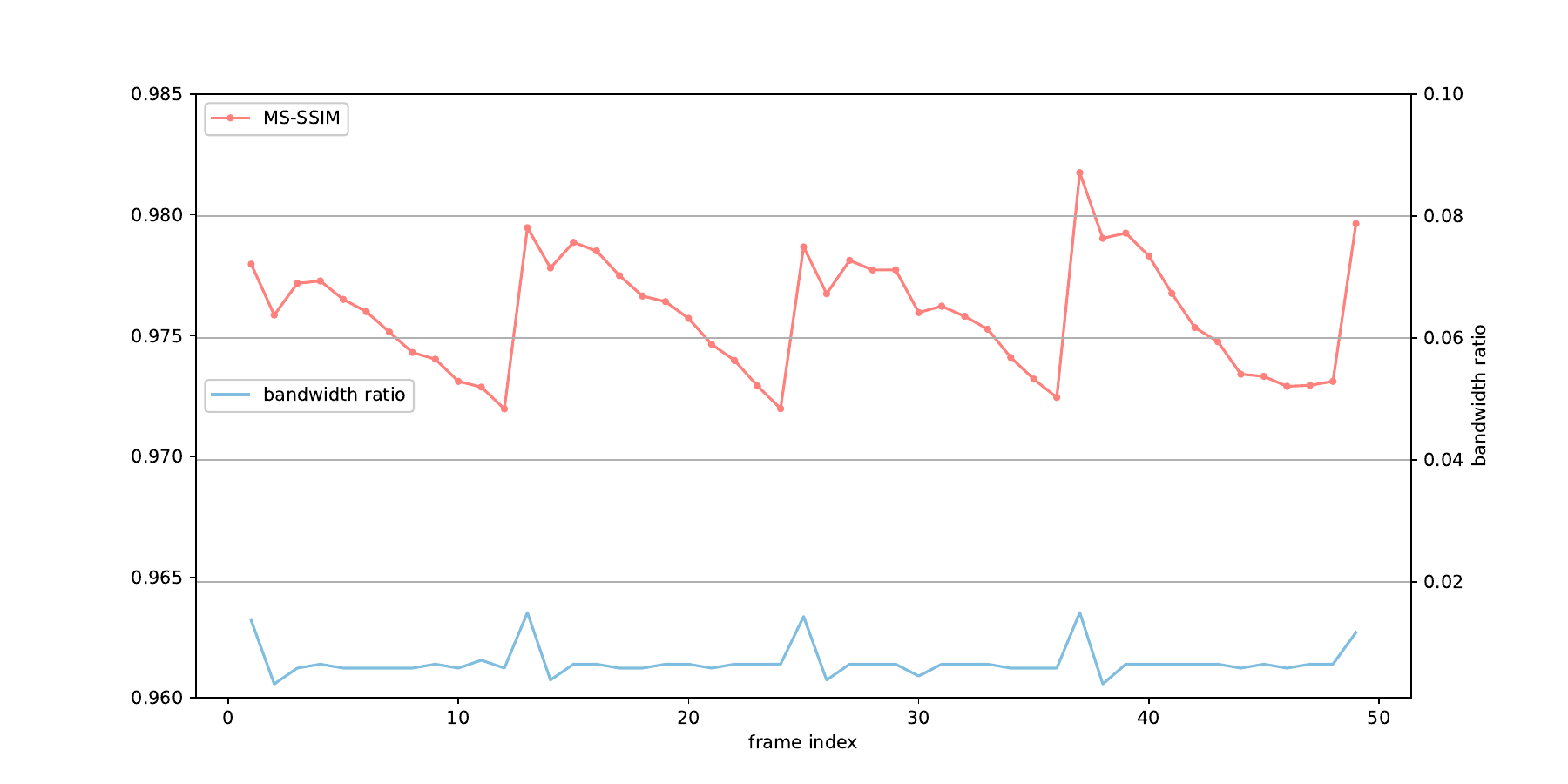}}
\caption{The bandwidth ratio and corresponding reconstruction performance curves of the test sequences, where (a) and (b) represent the results of sequence \emph{BQTerrace} at CSNR = 10dB, while (c) and (d) represent the results of \emph{videoSRC16} at CSNR = 5dB.}
\label{fig: 50framepsnr}
\end{figure*}

\begin{figure*}[hbtp]
\centering
\subfigure[]{
\includegraphics[width=0.46\linewidth]{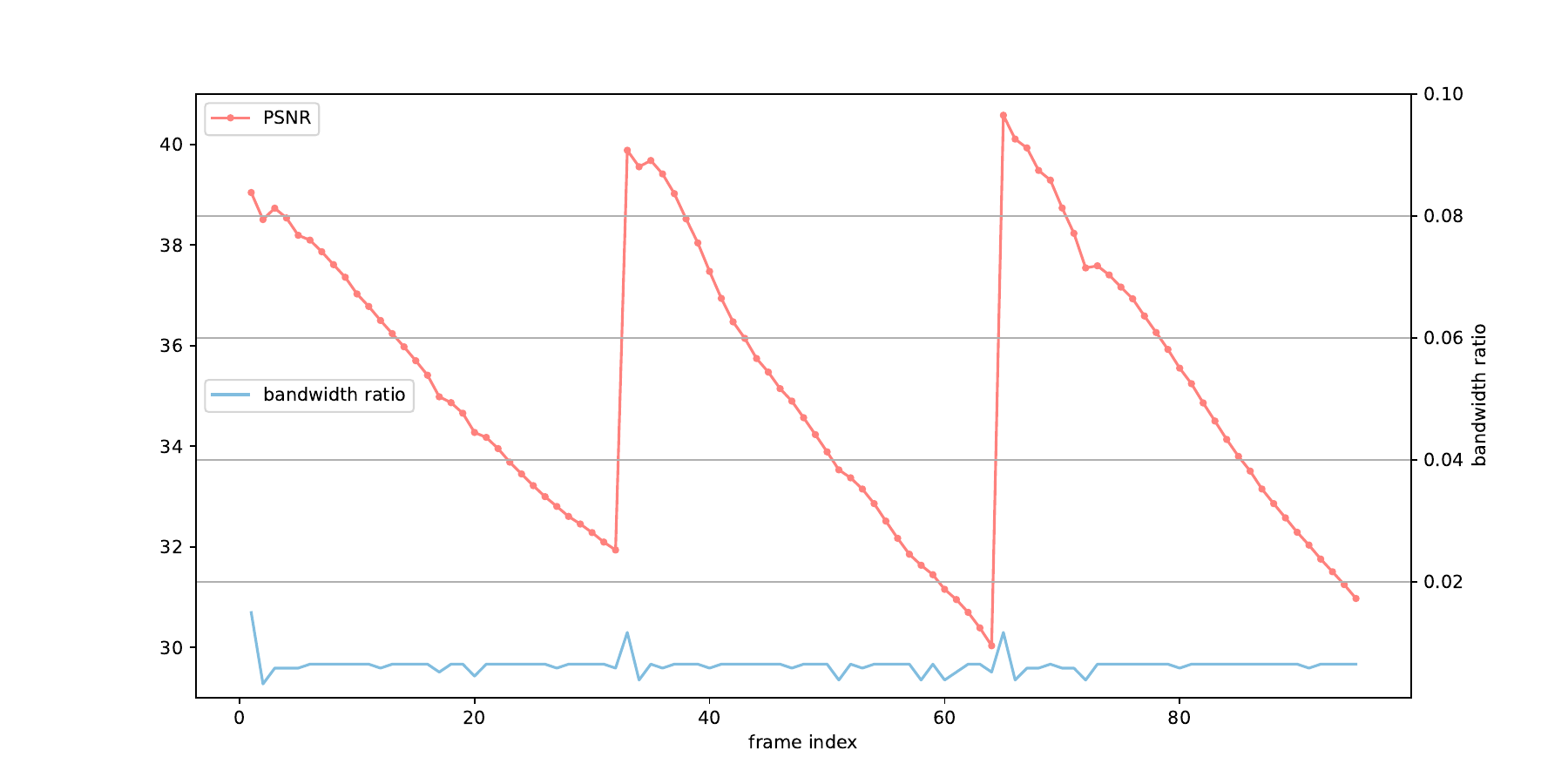}}
\hspace{-0.02\linewidth}
\subfigure[]{
\includegraphics[width=0.46\linewidth]{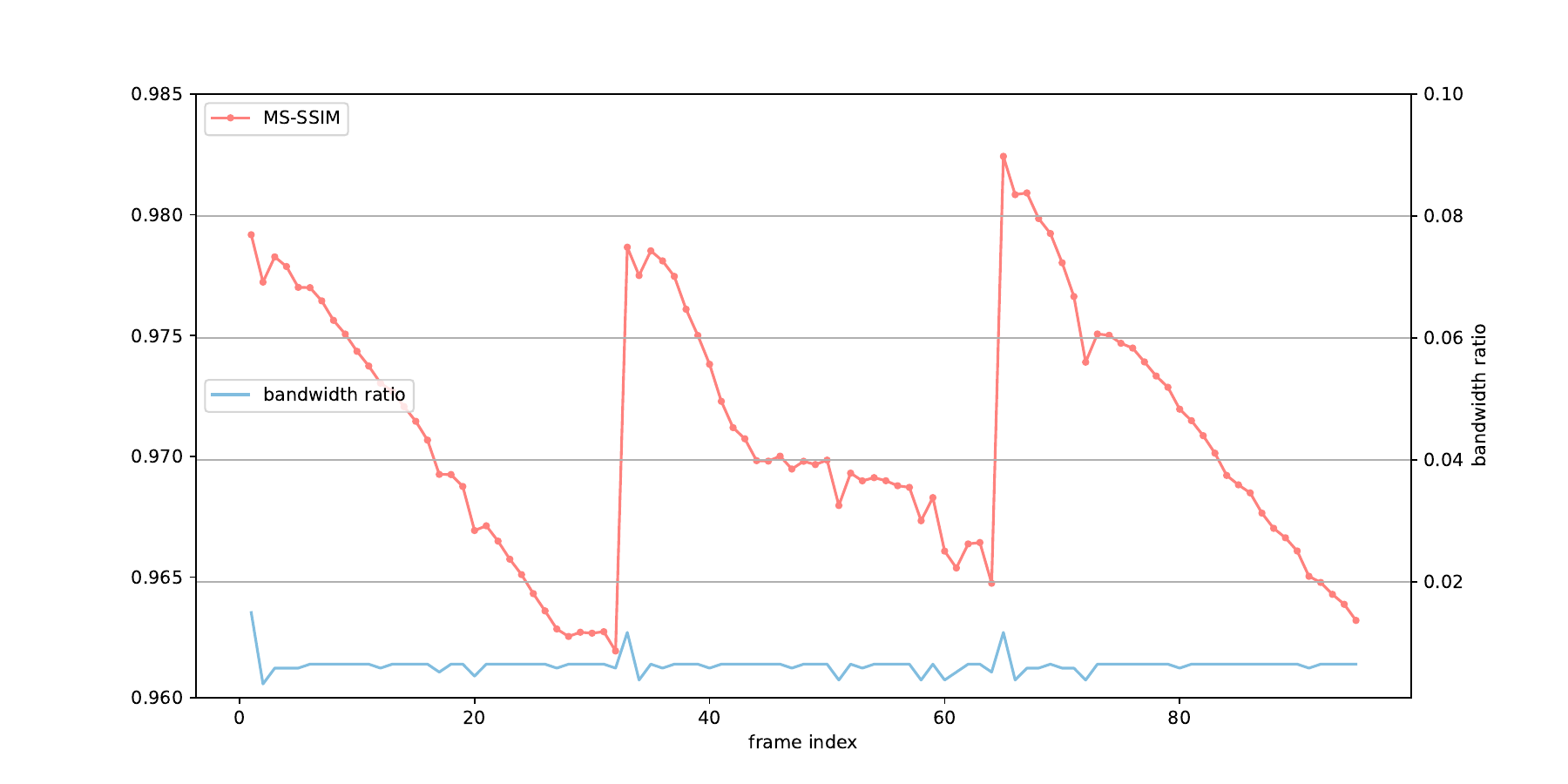}}
\caption{The bandwidth ratio and corresponding reconstruction performance curves of the test sequences, where (a) and (b) represent the results of represent the results of \emph{videoSRC16} at CSNR = 5dB.}
\label{fig: 96framepsnr}
\end{figure*}

For most video compression algorithms, error accumulation is an inevitable issue. Moreover, in communication systems based on pseudo-analog transmission, this problem is even more pronounced. This is because such methods inevitably require learning inter-frame correlations from asymmetric contexts, which is more challenging compared to learn from symmetric context as in DCVC \cite{dcvc}. 
DVST \cite{dvst} rely on using smaller GOP to mitigate this problem by frequent use of intra-frame coding modes to alleviate error accumulation. 
However, approximately 3dB of performance degradation still occurs at a GOP of 4.
However, smaller intra-frame coding periods reduce compression efficiency. Since testing DVST \cite{dvst} with longer GOP resulted in significant performance degradation and error accumulation, we experimented only with the separated scheme under a longer GOP. Fig. \ref{fig: gop12psnr} illustrates the R-D performance curves of our scheme tested under a longer GOP. We tested the GOP settings of HEVC dataset at 10 and MCL-JCV dataset at 12, following the settings in \cite{dvc6}. For most test videos, our schemes consistently outperforms H.265-based separable communication systems, demonstrating the stable performance of our method even under long GOP sequences. Fig. \ref{fig: gop12ssim} compares the MS-SSIM performance of various approaches under the same conditions. It can be observed that our method exhibits a more significant improvement in performance compared to non-neural network-based traditional methods in terms of this metric.

\newcolumntype{C}[1]{>{\centering\arraybackslash}p{#1}}
\newcommand{\imgcell}[2]{\makecell{\includegraphics[width=0.16\textwidth]{#1}\\\vphantom{Bpp}#2}}
\newcommand{\imgfullcell}[2]{\makecell{\includegraphics[width=0.22\textwidth]{#1}\\#2}}
\newcommand{\midrotate}[1]{\raisebox{0.8em}{\rotatebox[origin=c]{90}{#1}}}

\begin{figure*}[!ht]
    \centering
    \renewcommand{\arraystretch}{1}
    \setlength{\tabcolsep}{2pt}
    \footnotesize
    \begin{tabular}{@{}l C{2.5cm} C{2.5cm} C{2.5cm} C{2.5cm} C{2.5cm}@{}}
        &
        \shortstack{(a) Origin} &
        \shortstack{(b) Proposed} &
        \shortstack{(c) H.264} &
        \shortstack{(d) H.265} \\
        
        \imgfullcell{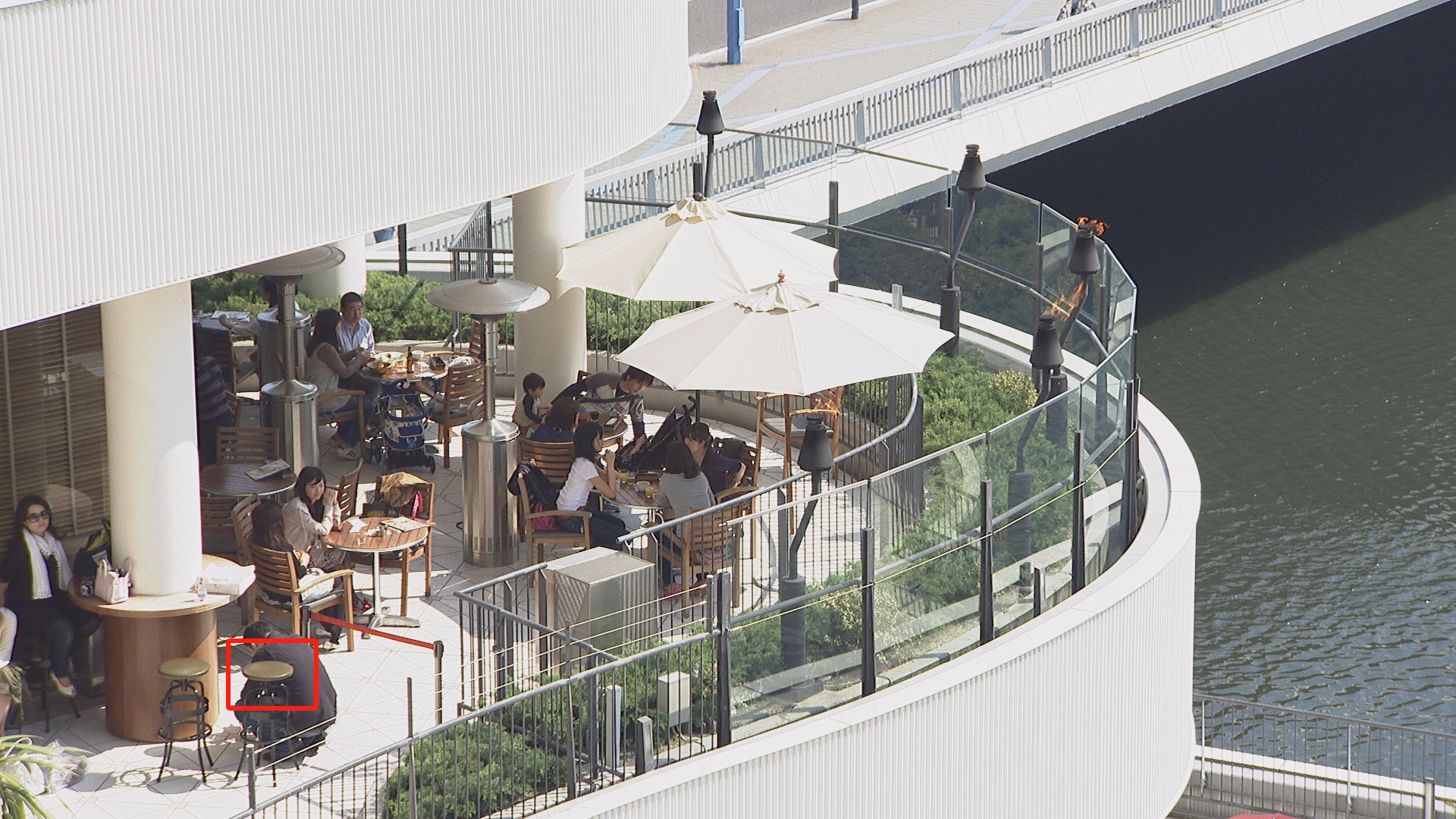}{1920×1080} &
        \imgcell{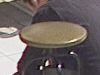}{} &
        \imgcell{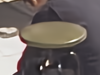}{CBR: 0.0208} &
        \imgcell{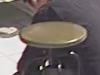}{CBR: 0.0236} &
        \imgcell{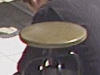}{CBR: 0.0241}
        \\
        
        \imgfullcell{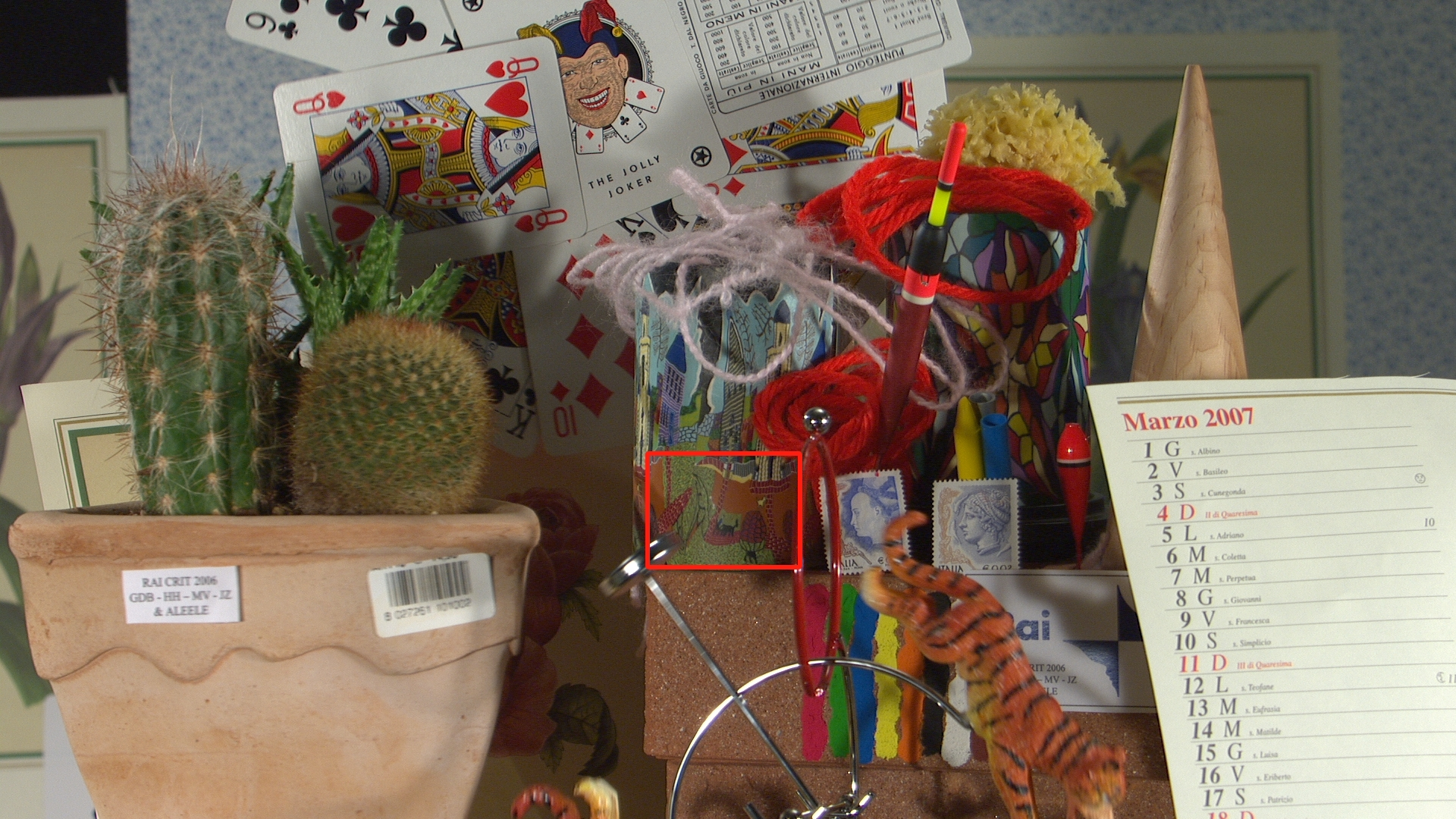}{1920×1080} &
        \imgcell{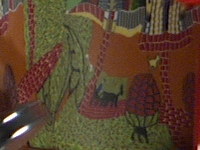}{} &
        \imgcell{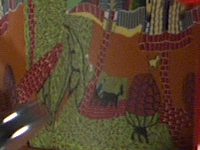}{CBR: 0.0182} &
        \imgcell{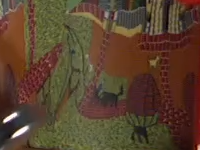}{CBR: 0.0241} &
        \imgcell{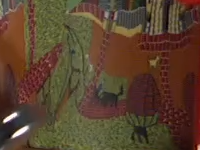}{CBR: 0.0225}
        \\
    \end{tabular}

    \caption{Examples of visual comparison. The first column shows the original frame. In the original frame, we mark the crop image with the red box. The least column show the reconstructed frames by using different transmission schemes over the AWGN channel at CSNR = 10dB.}
    \label{fig: vis}
\end{figure*}

\newcommand{\newimgcell}[2]{\makecell{\vspace{1pt}\includegraphics[height=2cm,keepaspectratio]{#1}\\#2}}
\newcommand{\newimgfullcell}[2]{\makecell{\vspace{1pt}\includegraphics[height=2cm,keepaspectratio]{#1}\\#2}}
\begin{figure*}[!ht]
    \centering
    \renewcommand{\arraystretch}{1}
    \setlength{\tabcolsep}{2pt}
    \footnotesize
    
    \renewcommand{\cellalign}{tc}
    \begin{tabular}[t]{@{}C{3.8cm} C{3.1cm} C{3.1cm} C{3.1cm}@{}}
        \shortstack{} &
        \shortstack{(a) CSNR = 5dB} &
        \shortstack{(b) CSNR = 10dB} &
        \shortstack{(c) CSNR = 15dB} \\
        
        \newimgfullcell{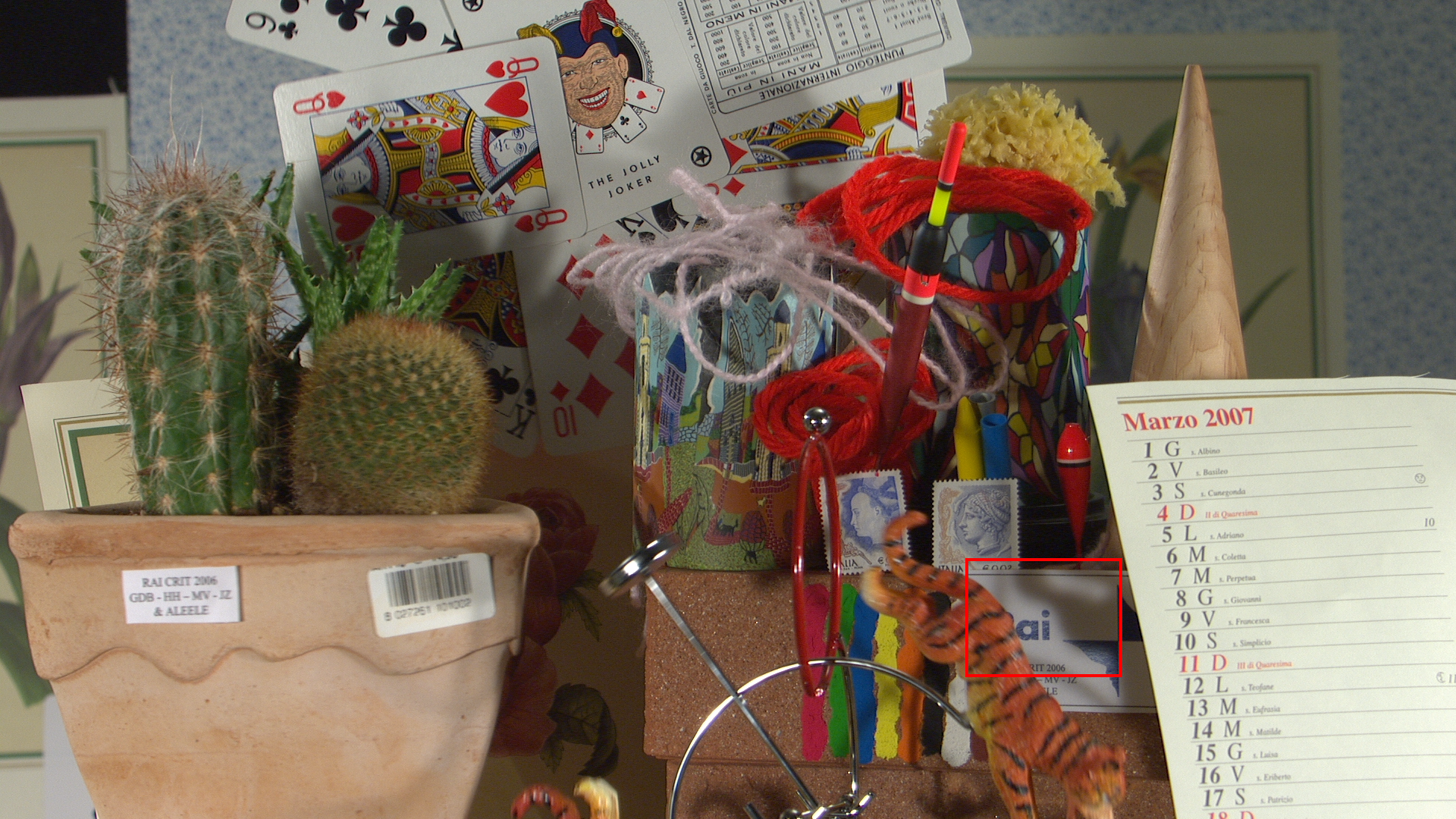}{Ground Truth} &
        \newimgcell{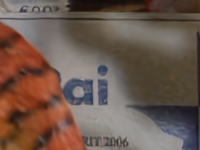}{Proposed (CSNR=5dB)} &
        \newimgcell{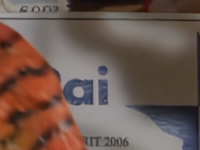}{Proposed (CSNR=10dB)} &
        \newimgcell{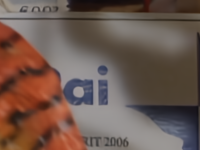}{Proposed (CSNR=15dB)} \\
        
        \shortstack{} &
        \shortstack{} &
        \shortstack{} &
        \shortstack{} \\
        
        \newimgcell{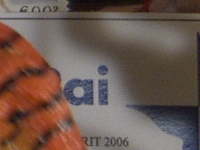}{Origin} &
        \newimgcell{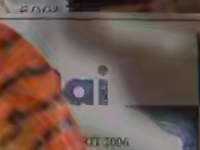}{H.265+2/3 LDPC\\+4QAM} &
        \newimgcell{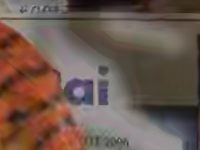}{H.265+2/3 LDPC\\+16QAM} &
        \newimgcell{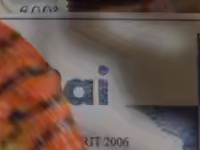}{H.265+1/2 LDPC\\+64QAM} \\
    \end{tabular}

    \caption{Visual comparison under different CSNRs. The first column shows the orginal frame and its cropped patch. In the original frame, we mark the crop image with the red box. The second to the fourth column shows the frames reconstructed by our proposed schemes or H.265 + LDPC over AWGN channel at various channel SNRs, respectively. The average bandwidth ratio of GOP is limited to 0.021.}
    \label{fig: vis2}
\end{figure*}

Beyond PSNR and MS-SSIM, we further evaluate perceptual quality using LPIPS \cite{lpips} metric. As shown in Fig. \ref{fig: gop12lpips}, our proposed scheme achieves consistently better perceptual quality across different bandwidth ratios on HEVC Class B and Class E compared with the separated scheme under a longer GOP. This indicates that our approach preserves finer perceptual details and generates fewer visible distortions, even under limited bandwidth. The consistent gap between our method and traditional digital-based transmission schemes demonstrates that the proposed content-adaptive variable bandwidth strategy not only enhances objective fidelity but also improves perceptual realism.

We can observe that our model encodes videos into channel symbols with different bandwidth compression ratios in different sequences. This is because our MEM can adaptively adjust the bandwidth ratio according to the content of the video. However, the fluctuations in bandwidth are not significant, thus ensuring the stability of the coding process. Additionally, we also demonstrate the R-D performance of our method on consecutive test video sequences. Figs. \ref{fig: 50framepsnr} and \ref{fig: 96framepsnr} illustrate the performance of our method on the first 50 and 96 frames of video sequences under GOP lengths of 10 and 32, respectively. In the curves of bandwidth compression ratio, we can observe noticeable spikes, which are caused by the fact that I-frames cannot refer to the context and therefore require more bandwidth to achieve performance similar to P-frames. P-frames, on the other hand, can achieve high compression ratios as they can refer to more context. The error accumulation is properly mitigated in terms of PSNR and MS-SSIM performance by guiding the network to learn coding from asymmetric contexts, although it still exists. In the long-GOP setting (GOP = 32), the variation between I-frame and P-frames becomes more pronounced due to the extended reference interval. Nevertheless, the overall PSNR and MS-SSIM curves remain stable, demonstrating that the proposed framework effectively suppresses error accumulation through asymmetric context modeling and feature propagation.


In Fig. \ref{fig: vis}, the visual effects of our proposed scheme on the test samples are demonstrated. It can be clearly observed that compared to other approaches, our method preserves more details during the image processing. Also, in Fig. \ref{fig: vis2}, under various channel SNRs, our proposed method maintains the structural integrity of image content and visual coherence. Even at low CSNR levels, it effectively suppresses blocking and ringing artifacts that are evident in the H.265 + LDPC + QAM scheme, resulting in smoother regions and fewer visible distortions. These results demonstrate the robustness of the proposed framework against channel degradation and its advantage in preserving perceptual quality.

\begin{figure}[!t]
\centering
\subfigure[]{
\includegraphics[width=0.45\linewidth]{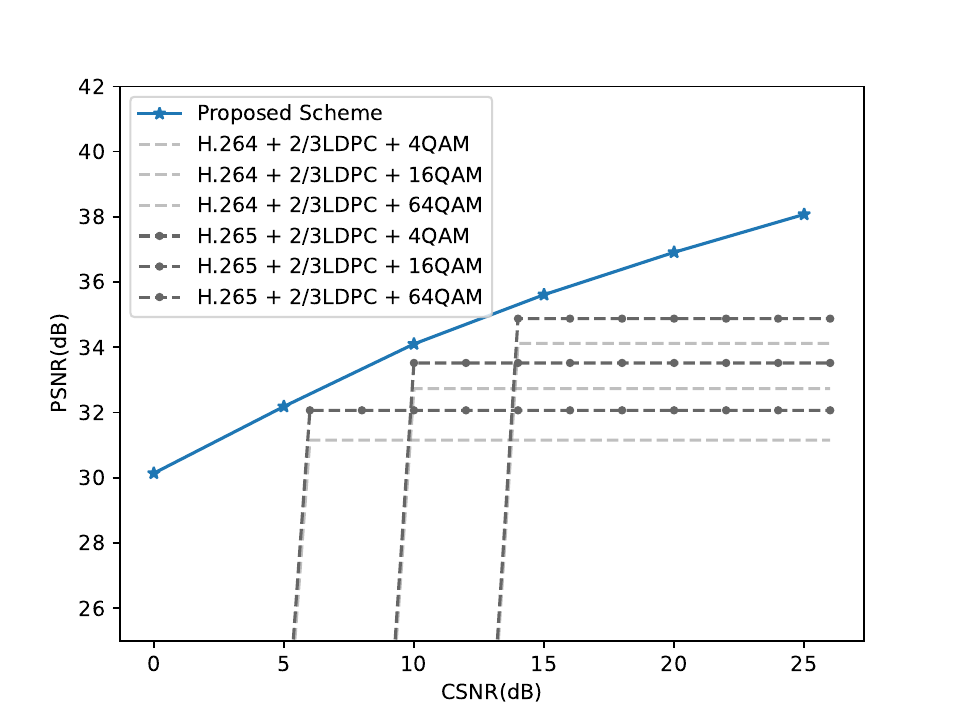}}
\subfigure[]{
\includegraphics[width=0.45\linewidth]{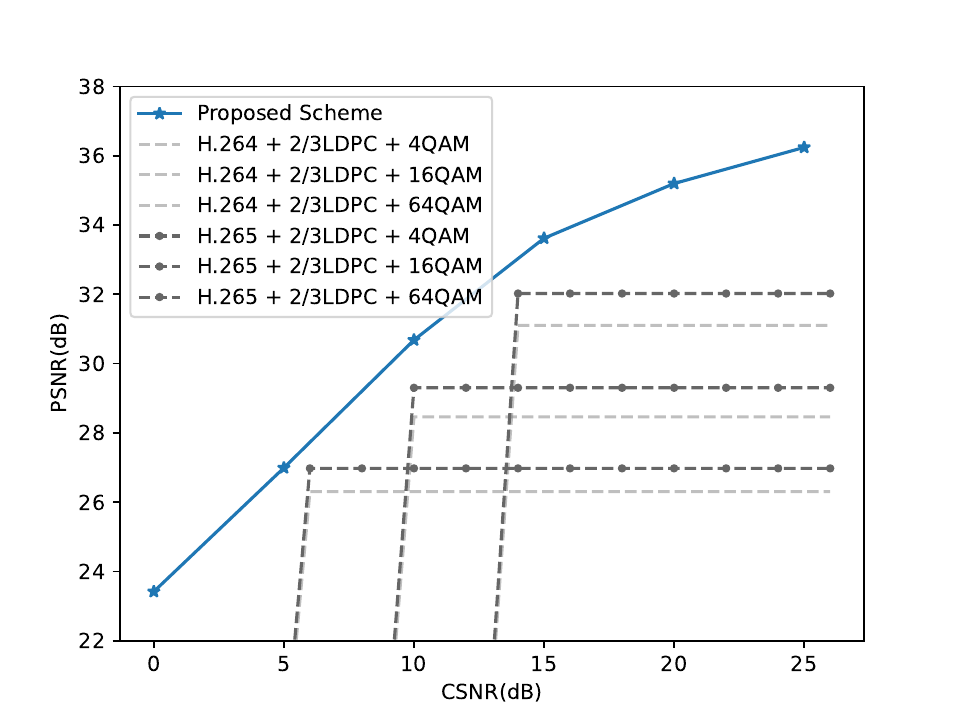}}
\caption{PSNR performance versus the CSNR of relevant schemes over the AWGN channel.}
\label{fig: csnr}
\end{figure}

\begin{figure}[!t]
\centering
\includegraphics[width=0.55\linewidth]{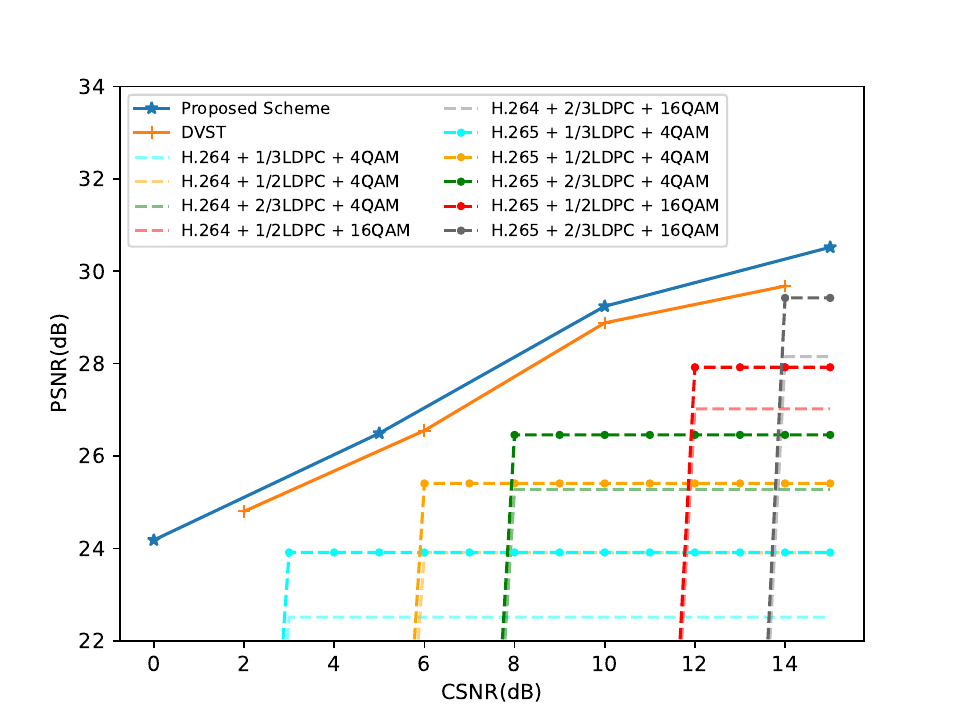}
\caption{PSNR performance versus the CSNR over the Rayleigh fading channel.}
\label{fig: rayleigh}
\end{figure}

We also tested the performance of the proposed scheme under various CSNRs to evaluate the channel adaptability of our scheme. To ensure a fair comparison, we adjusted the bandwidth of the comparison schemes accordingly to ensure that the transmission bandwidths are as close as possible to each other. In order to make the bandwidths at different CSNRs as close as possible, we set six CSNR $= {0, 5, 10, 15, 20, 25}$dB with $\lambda$ = (3e-3, 2e-3, 1e-3, 8e-4, 6e-4, 2e-4). Fig. \ref{fig: csnr} shows the performance curves of PSNR as the CSNR varies. For the comparison schemes, we used combinations of 2/3 LDPC and 4QAM, 16QAM, and 64QAM. The blue curve represents the performance of the proposed scheme trained and tested under various CSNRs. We can observe that our approach exhibits improved performance corresponding to the increase in CSNR, demonstrating excellent channel robustness.

Furthermore, to further validate the robustness of the proposed scheme under more practical fading conditions, we evaluate its performance over the Rayleigh fading channel across different CSNR levels on HEVC Class D test sequence. In practice, our models of the Rayleigh fading channel are fine-tuned from baseline models trained under the AWGN channel at the same CSNR. As shown in Fig. \ref{fig: rayleigh}, our proposed scheme maintains a clear advantage across all CSNR levels, demonstrating strong robustness and adaptability to fading environments.

\begin{figure}[!t]
\centering
\includegraphics[width=0.45\linewidth]{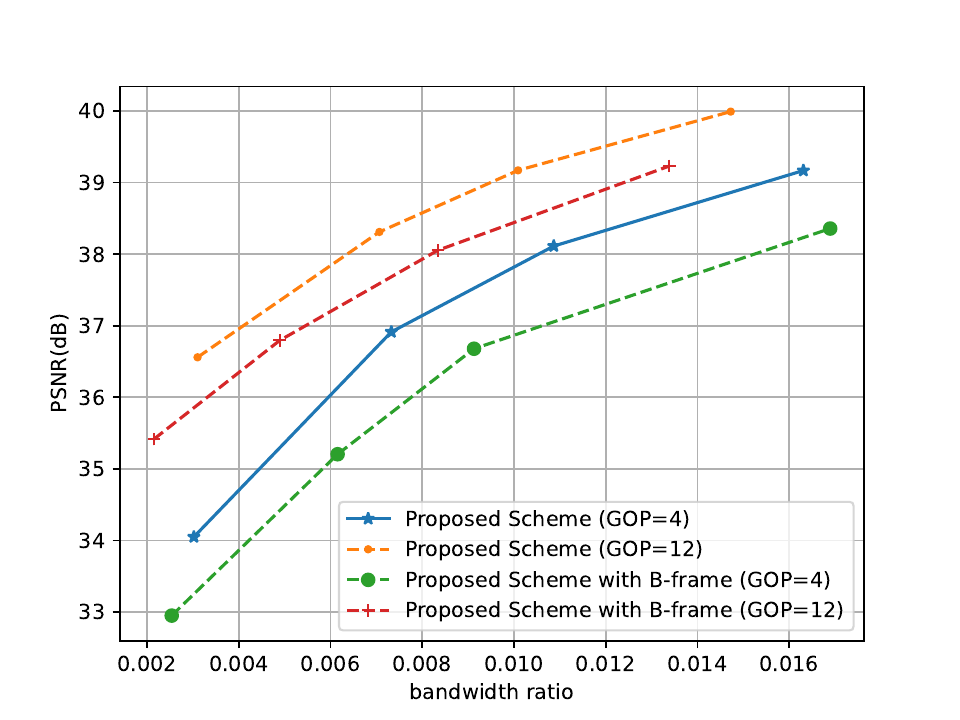}
\caption{PSNR versus the average bandwidth ratio for GOP structures with and without B-frames under GOP=4 and GOP=12 on HEVC Class E test sequence.}
\label{fig: bframe}
\end{figure}

To evaluate the robustness of our method under more general random-access service mode, we further consider the GOP structures including B-frames. In the GOP structures including B-frames, we insert one B-frame between every two reference frames by default. As shown in Fig. \ref{fig: bframe}, introducing B-frames leads to moderate performance degradation compared with the I/P-frame configuration. This degradation mainly arises from error accumulation: encoding a B-frame requires both its preceding and succeeding reference frames, so distortions in either reference can compound during bidirectional prediction. Nevertheless, the results indicate that our proposed method remains competitive performance and can be extended to more general GOP structures.

\renewcommand{\arraystretch}{1.5}

\begin{table}[!t]
\caption{Performance influence of different components of our proposed scheme over the AWGN channel with CSNR = 10dB.}
\label{table1}
\centering
\begin{tabular*}{\textwidth}{@{\extracolsep\fill}ccccccc}
\toprule
+MEM & +FP & +FR Loss & HEVC B & HEVC C & HEVC E & MCL-JCV\\
\midrule
\checkmark & \checkmark & \checkmark & 0.0 & 0.0 & 0.0 & 0.0 \\
$\times$  & \checkmark & \checkmark & -2.3\% & -4.8\% & -2.4\% & -3.3\% \\
\checkmark & \checkmark & $\times$ & -3.7\% & -5.9\% & -2.9\% & -5.4\% \\
\checkmark & $\times$ & $\times$  & -5.1\% & -9.5\% & -4.3\% & -7.0\% \\
$\times$  & $\times$  & $\times$  & -7.8\% & -13.0\% & -6.8\% & -11.8\% \\
\botrule
\end{tabular*}
\end{table}

To further analyze the effectiveness of each component in our proposed framework, we conducted an ablation study on the Mask-based Entropy Module (MEM), the Feature Propagation mechanism (FP), and the Feature Reconstruction Loss (FR Loss), as summarized in Table \ref{table1}. Table \ref{table1} reports the PSNR degradation percentages relative to the complete configuration on multiple datasets. The baseline is the full configuration of our proposed scheme. When $MEM$ is removed, the number of latent channels used for transmission is fixed to 32 channels to maintain a constant channel bandwidth ratio. In contrast, when the MEM is applied, the bandwidth ratio is controlled by adjusting the parameter $\lambda$ to ensure comparable transmission rates across different component combinations within the same dataset. As shown in Table \ref{table1}, removing either the MEM or the FP leads to a notable performance degradation, highlighting their crucial roles in adaptive bandwidth control and mitigating error accumulation under long-GOP settings. The FR Loss enhances the use of reference information from previous frames, further improving overall reconstruction quality. These results confirm that the proposed modules jointly enhance both coding efficiency and performance in our proposed DeepJSCC framework.


\renewcommand{\arraystretch}{1.5}

\begin{table}[!t]
\label{table2}
\caption{Calculation and storage comparison between DVST and our method.}\label{tab2}
\begin{tabular*}{\textwidth}{@{\extracolsep\fill}ccccc}
\toprule%
& \multicolumn{2}{@{}c@{}}{I-frame(Batch=32)} & \multicolumn{2}{@{}c@{}}{P-frame(Batch=2)} \\\cmidrule{2-3}\cmidrule{4-5}%
Method & FLOPS & Params & FLOPS & Params\\
\midrule
DVST  & 2048.97G & 27.56M  & 484.60G & \pmb{13.65M} \\
ours  & \pmb{902.66G} & \pmb{17.82M} & \pmb{277.43G} & 15.47M \\
\botrule
\end{tabular*}
\end{table}

Finally, we calculate the Floating-point Operations Per Second (FLOPS) and parameters required for DVST \cite{dvst} and proposed method. To provide a fair comparison, we present the FLOPS and parameters for the I-frame and P-frame networks of each method in Table \ref{table2}. We follow the training settings to set the batch size for I-frame and P-frame networks. 
It is evident that, in the I-frame network, the FLOPS of DVST \cite{dvst} is more than double the FLOPS of our method and parameters of our method is about 35\% less than DVST \cite{dvst}. 
Then, in the P-frame network, our method demands slightly more parameters than DVST \cite{dvst}. But the amount of floating-point computation required is greatly reduced, by about 42\% compared to DVST \cite{dvst}.

\section{Conclusion}
In this paper, we propose a wireless video transmission scheme based on asymmetric context. We implicitly learn to mine the correlation between video frames from asymmetric contexts via a neural network. Moreover we introduce feature propagation to exploit multi-frame correlation and reduce error accumulation by propagating intermediate features of the network independently at the encoder and decoder. Finally, we propose to utilize the entropy model and masking mechanism to achieve variable bandwidth transmission based on video content. Experimental results show that our scheme achieves better performance compared to state-of-the-art schemes. In the future, we will extend the proposed scheme to more channel models and introduce the channel attention mechanism to enhance the channel adaptability.

\bmhead{Acknowledgements}

This work was supported in part by the Joint Funds for Railway Fundamental Research of National Natural Science Foundation of China under Grant No.U2368201 and the Changsha Natural Science Foundation under Grant No.kq2502118.


\bibliography{ref}


\begin{thebibliography}{56}
\ifx \bisbn   \undefined \def \bisbn  #1{ISBN #1}\fi
\ifx \binits  \undefined \def \binits#1{#1}\fi
\ifx \bauthor  \undefined \def \bauthor#1{#1}\fi
\ifx \batitle  \undefined \def \batitle#1{#1}\fi
\ifx \bjtitle  \undefined \def \bjtitle#1{#1}\fi
\ifx \bvolume  \undefined \def \bvolume#1{\textbf{#1}}\fi
\ifx \byear  \undefined \def \byear#1{#1}\fi
\ifx \bissue  \undefined \def \bissue#1{#1}\fi
\ifx \bfpage  \undefined \def \bfpage#1{#1}\fi
\ifx \blpage  \undefined \def \blpage #1{#1}\fi
\ifx \burl  \undefined \def \burl#1{\textsf{#1}}\fi
\ifx \doiurl  \undefined \def \doiurl#1{\url{https://doi.org/#1}}\fi
\ifx \betal  \undefined \def \betal{\textit{et al.}}\fi
\ifx \binstitute  \undefined \def \binstitute#1{#1}\fi
\ifx \binstitutionaled  \undefined \def \binstitutionaled#1{#1}\fi
\ifx \bctitle  \undefined \def \bctitle#1{#1}\fi
\ifx \beditor  \undefined \def \beditor#1{#1}\fi
\ifx \bpublisher  \undefined \def \bpublisher#1{#1}\fi
\ifx \bbtitle  \undefined \def \bbtitle#1{#1}\fi
\ifx \bedition  \undefined \def \bedition#1{#1}\fi
\ifx \bseriesno  \undefined \def \bseriesno#1{#1}\fi
\ifx \blocation  \undefined \def \blocation#1{#1}\fi
\ifx \bsertitle  \undefined \def \bsertitle#1{#1}\fi
\ifx \bsnm \undefined \def \bsnm#1{#1}\fi
\ifx \bsuffix \undefined \def \bsuffix#1{#1}\fi
\ifx \bparticle \undefined \def \bparticle#1{#1}\fi
\ifx \barticle \undefined \def \barticle#1{#1}\fi
\bibcommenthead
\ifx \bconfdate \undefined \def \bconfdate #1{#1}\fi
\ifx \botherref \undefined \def \botherref #1{#1}\fi
\ifx \url \undefined \def \url#1{\textsf{#1}}\fi
\ifx \bchapter \undefined \def \bchapter#1{#1}\fi
\ifx \bbook \undefined \def \bbook#1{#1}\fi
\ifx \bcomment \undefined \def \bcomment#1{#1}\fi
\ifx \oauthor \undefined \def \oauthor#1{#1}\fi
\ifx \citeauthoryear \undefined \def \citeauthoryear#1{#1}\fi
\ifx \endbibitem  \undefined \def \endbibitem {}\fi
\ifx \bconflocation  \undefined \def \bconflocation#1{#1}\fi
\ifx \arxivurl  \undefined \def \arxivurl#1{\textsf{#1}}\fi
\csname PreBibitemsHook\endcsname

\bibitem[\protect\citeauthoryear{Index}{2017}]{cisco}
\begin{botherref}
\oauthor{\bsnm{Index}, \binits{C.V.N.}}:
Forecast and methodology, 2016--2021
(2017)
\end{botherref}
\endbibitem

\bibitem[\protect\citeauthoryear{Fresia et~al.}{2010}]{jscc1}
\begin{barticle}
\bauthor{\bsnm{Fresia}, \binits{M.}},
\bauthor{\bsnm{Perez-Cruz}, \binits{F.}},
\bauthor{\bsnm{Poor}, \binits{H.V.}},
\bauthor{\bsnm{Verdu}, \binits{S.}}:
\batitle{Joint source and channel coding}.
\bjtitle{IEEE Signal Processing Magazine}
\bvolume{27}(\bissue{6}),
\bfpage{104}--\blpage{113}
(\byear{2010})
\doiurl{10.1109/MSP.2010.938080}
\end{barticle}
\endbibitem

\bibitem[\protect\citeauthoryear{Sayood et~al.}{2000}]{jscc2}
\begin{barticle}
\bauthor{\bsnm{Sayood}, \binits{K.}},
\bauthor{\bsnm{Otu}, \binits{H.H.}},
\bauthor{\bsnm{Demir}, \binits{N.}}:
\batitle{Joint source/channel coding for variable length codes}.
\bjtitle{IEEE Transactions on Communications}
\bvolume{48}(\bissue{5}),
\bfpage{787}--\blpage{794}
(\byear{2000})
\doiurl{10.1109/26.843191}
\end{barticle}
\endbibitem

\bibitem[\protect\citeauthoryear{Gallager}{1968}]{jscc3}
\begin{bbook}
\bauthor{\bsnm{Gallager}, \binits{R.G.}}:
\bbtitle{Information Theory and Reliable Communication}.
\bpublisher{Wiley},
\blocation{Hoboken, NJ, USA}
(\byear{1968})
\end{bbook}
\endbibitem

\bibitem[\protect\citeauthoryear{Kostina and Verd{\'u}}{2013}]{jscc4}
\begin{barticle}
\bauthor{\bsnm{Kostina}, \binits{V.}},
\bauthor{\bsnm{Verd{\'u}}, \binits{S.}}:
\batitle{Lossy joint source-channel coding in the finite blocklength regime}.
\bjtitle{IEEE Transactions on Information Theory}
\bvolume{59}(\bissue{5}),
\bfpage{2545}--\blpage{2575}
(\byear{2013})
\doiurl{10.1109/TIT.2013.2238657}
\end{barticle}
\endbibitem

\bibitem[\protect\citeauthoryear{Goblick}{1969}]{jscc5}
\begin{barticle}
\bauthor{\bsnm{Goblick}, \binits{T.}}:
\batitle{A coding theorem for time-discrete analog data sources}.
\bjtitle{IEEE Transactions on Information Theory}
\bvolume{15}(\bissue{3}),
\bfpage{401}--\blpage{407}
(\byear{1969})
\doiurl{10.1109/TIT.1969.1054303}
\end{barticle}
\endbibitem

\bibitem[\protect\citeauthoryear{Gastpar et~al.}{2003}]{jscc6}
\begin{barticle}
\bauthor{\bsnm{Gastpar}, \binits{M.}},
\bauthor{\bsnm{Rimoldi}, \binits{B.}},
\bauthor{\bsnm{Vetterli}, \binits{M.}}:
\batitle{To code, or not to code: Lossy source-channel communication
  revisited}.
\bjtitle{IEEE Transactions on Information Theory}
\bvolume{49}(\bissue{5}),
\bfpage{1147}--\blpage{1158}
(\byear{2003})
\doiurl{10.1109/TIT.2003.810631}
\end{barticle}
\endbibitem

\bibitem[\protect\citeauthoryear{Bourtsoulatze et~al.}{2019}]{djscc1}
\begin{barticle}
\bauthor{\bsnm{Bourtsoulatze}, \binits{E.}},
\bauthor{\bsnm{Kurka}, \binits{D.B.}},
\bauthor{\bsnm{G{\"u}nd{\"u}z}, \binits{D.}}:
\batitle{Deep joint source-channel coding for wireless image transmission}.
\bjtitle{IEEE Transactions on Cognitive Communications and Networking}
\bvolume{5}(\bissue{3}),
\bfpage{567}--\blpage{579}
(\byear{2019})
\doiurl{10.1109/TCCN.2019.2919300}
\end{barticle}
\endbibitem

\bibitem[\protect\citeauthoryear{Kurka and G{\"u}nd{\"u}z}{2020}]{djscc2}
\begin{barticle}
\bauthor{\bsnm{Kurka}, \binits{D.B.}},
\bauthor{\bsnm{G{\"u}nd{\"u}z}, \binits{D.}}:
\batitle{Deepjscc-f: Deep joint source-channel coding of images with feedback}.
\bjtitle{IEEE Journal on Selected Areas in Information Theory}
\bvolume{1}(\bissue{1}),
\bfpage{178}--\blpage{193}
(\byear{2020})
\doiurl{10.1109/JSAIT.2020.2987203}
\end{barticle}
\endbibitem

\bibitem[\protect\citeauthoryear{Ding et~al.}{2021}]{djscc3}
\begin{bchapter}
\bauthor{\bsnm{Ding}, \binits{M.}},
\bauthor{\bsnm{Li}, \binits{J.}},
\bauthor{\bsnm{Ma}, \binits{M.}},
\bauthor{\bsnm{Fan}, \binits{X.}}:
\bctitle{Snr-adaptive deep joint source-channel coding for wireless image
  transmission}.
In: \bbtitle{Proceedings of the IEEE International Conference on Acoustics,
  Speech, and Signal Processing (ICASSP)},
pp. \bfpage{1555}--\blpage{1559}
(\byear{2021}).
\doiurl{10.1109/ICASSP39728.2021.9414037}
\end{bchapter}
\endbibitem

\bibitem[\protect\citeauthoryear{Yang and Kim}{2022}]{djscc4}
\begin{bchapter}
\bauthor{\bsnm{Yang}, \binits{M.}},
\bauthor{\bsnm{Kim}, \binits{H.-S.}}:
\bctitle{Deep joint source-channel coding for wireless image transmission with
  adaptive rate control}.
In: \bbtitle{Proceedings of the IEEE International Conference on Acoustics,
  Speech, and Signal Processing (ICASSP)},
pp. \bfpage{5193}--\blpage{5197}
(\byear{2022}).
\doiurl{10.1109/ICASSP43922.2022.9746335}
\end{bchapter}
\endbibitem

\bibitem[\protect\citeauthoryear{Kurka and G{\"u}nd{\"u}z}{2021}]{djscc5}
\begin{barticle}
\bauthor{\bsnm{Kurka}, \binits{D.B.}},
\bauthor{\bsnm{G{\"u}nd{\"u}z}, \binits{D.}}:
\batitle{Bandwidth-agile image transmission with deep joint source-channel
  coding}.
\bjtitle{IEEE Transactions on Wireless Communications}
\bvolume{20}(\bissue{12}),
\bfpage{8081}--\blpage{8095}
(\byear{2021})
\doiurl{10.1109/TWC.2021.3090048}
\end{barticle}
\endbibitem

\bibitem[\protect\citeauthoryear{Jankowski et~al.}{2020}]{djscc6}
\begin{bchapter}
\bauthor{\bsnm{Jankowski}, \binits{M.}},
\bauthor{\bsnm{G{\"u}nd{\"u}z}, \binits{D.}},
\bauthor{\bsnm{Mikolajczyk}, \binits{K.}}:
\bctitle{Deep joint source-channel coding for wireless image retrieval}.
In: \bbtitle{Proceedings of the IEEE International Conference on Acoustics,
  Speech, and Signal Processing (ICASSP)},
pp. \bfpage{5070}--\blpage{5074}
(\byear{2020}).
\doiurl{10.1109/ICASSP40776.2020.9054078}
\end{bchapter}
\endbibitem

\bibitem[\protect\citeauthoryear{Tung et~al.}{2022}]{djscc7}
\begin{barticle}
\bauthor{\bsnm{Tung}, \binits{T.-Y.}},
\bauthor{\bsnm{Kurka}, \binits{D.B.}},
\bauthor{\bsnm{Jankowski}, \binits{M.}},
\bauthor{\bsnm{G{\"u}nd{\"u}z}, \binits{D.}}:
\batitle{Deepjscc-q: Constellation constrained deep joint source-channel
  coding}.
\bjtitle{IEEE Journal on Selected Areas in Information Theory}
\bvolume{3}(\bissue{4}),
\bfpage{720}--\blpage{731}
(\byear{2022})
\doiurl{10.1109/JSAIT.2022.3231042}
\end{barticle}
\endbibitem

\bibitem[\protect\citeauthoryear{Sun et~al.}{2021}]{djscc8}
\begin{barticle}
\bauthor{\bsnm{Sun}, \binits{S.}},
\bauthor{\bsnm{He}, \binits{T.}},
\bauthor{\bsnm{Chen}, \binits{Z.}}:
\batitle{Semantic structured image coding framework for multiple intelligent
  applications}.
\bjtitle{IEEE Transactions on Circuits and Systems for Video Technology}
\bvolume{31}(\bissue{9}),
\bfpage{3631}--\blpage{3642}
(\byear{2021})
\doiurl{10.1109/TCSVT.2020.3042517}
\end{barticle}
\endbibitem

\bibitem[\protect\citeauthoryear{Xu et~al.}{2022}]{djscc9}
\begin{barticle}
\bauthor{\bsnm{Xu}, \binits{J.}},
\bauthor{\bsnm{Ai}, \binits{B.}},
\bauthor{\bsnm{Chen}, \binits{W.}},
\bauthor{\bsnm{Yang}, \binits{A.}},
\bauthor{\bsnm{Sun}, \binits{P.}},
\bauthor{\bsnm{Rodrigues}, \binits{M.}}:
\batitle{Wireless image transmission using deep source channel coding with
  attention modules}.
\bjtitle{IEEE Transactions on Circuits and Systems for Video Technology}
\bvolume{32}(\bissue{4}),
\bfpage{2315}--\blpage{2328}
(\byear{2022})
\doiurl{10.1109/TCSVT.2021.3082521}
\end{barticle}
\endbibitem

\bibitem[\protect\citeauthoryear{Huang et~al.}{2021}]{djscc10}
\begin{bchapter}
\bauthor{\bsnm{Huang}, \binits{D.}},
\bauthor{\bsnm{Tao}, \binits{X.}},
\bauthor{\bsnm{Gao}, \binits{F.}},
\bauthor{\bsnm{Lu}, \binits{J.}}:
\bctitle{Deep learning-based image semantic coding for semantic
  communications}.
In: \bbtitle{Proceedings of the IEEE Global Communications Conference
  (GLOBECOM)},
pp. \bfpage{1}--\blpage{6}
(\byear{2021}).
\doiurl{10.1109/GLOBECOM46510.2021.9685667}
\end{bchapter}
\endbibitem

\bibitem[\protect\citeauthoryear{Li et~al.}{2021}]{dcvc}
\begin{bchapter}
\bauthor{\bsnm{Li}, \binits{J.}},
\bauthor{\bsnm{Li}, \binits{B.}},
\bauthor{\bsnm{Lu}, \binits{Y.}}:
\bctitle{Deep contextual video compression},
vol. \bseriesno{34},
pp. \bfpage{18114}--\blpage{18125}
(\byear{2021})
\end{bchapter}
\endbibitem

\bibitem[\protect\citeauthoryear{Li et~al.}{2023}]{dcvc-dc}
\begin{bchapter}
\bauthor{\bsnm{Li}, \binits{J.}},
\bauthor{\bsnm{Li}, \binits{B.}},
\bauthor{\bsnm{Lu}, \binits{Y.}}:
\bctitle{Neural video compression with diverse contexts}.
In: \bbtitle{Proceedings of the IEEE Conference on Computer Vision and Pattern
  Recognition (CVPR)},
pp. \bfpage{22616}--\blpage{22626}
(\byear{2023}).
\doiurl{10.1109/CVPR52729.2023.02166}
\end{bchapter}
\endbibitem

\bibitem[\protect\citeauthoryear{Wang et~al.}{2022}]{dvst}
\begin{barticle}
\bauthor{\bsnm{Wang}, \binits{S.}},
\bauthor{\bsnm{Dai}, \binits{J.}},
\bauthor{\bsnm{Liang}, \binits{Z.}},
\bauthor{\bsnm{Niu}, \binits{K.}},
\bauthor{\bsnm{Si}, \binits{Z.}},
\bauthor{\bsnm{Dong}, \binits{C.}},
\bauthor{\bsnm{Qin}, \binits{X.}},
\bauthor{\bsnm{Zhang}, \binits{P.}}:
\batitle{Wireless deep video semantic transmission}.
\bjtitle{IEEE Journal on Selected Areas in Communications}
\bvolume{41}(\bissue{1}),
\bfpage{214}--\blpage{229}
(\byear{2022})
\doiurl{10.1109/JSAC.2022.3221977}
\end{barticle}
\endbibitem

\bibitem[\protect\citeauthoryear{Lu et~al.}{2019}]{dvc6}
\begin{bchapter}
\bauthor{\bsnm{Lu}, \binits{G.}},
\bauthor{\bsnm{Ouyang}, \binits{W.}},
\bauthor{\bsnm{Xu}, \binits{D.}},
\bauthor{\bsnm{Zhang}, \binits{X.}},
\bauthor{\bsnm{Cai}, \binits{C.}},
\bauthor{\bsnm{Gao}, \binits{Z.}}:
\bctitle{Dvc: An end-to-end deep video compression framework}.
In: \bbtitle{Proceedings of the IEEE Conference on Computer Vision and Pattern
  Recognition (CVPR)},
pp. \bfpage{10998}--\blpage{11007}
(\byear{2019}).
\doiurl{10.1109/CVPR.2019.01126}
\end{bchapter}
\endbibitem

\bibitem[\protect\citeauthoryear{Hu et~al.}{2021}]{dvc1}
\begin{bchapter}
\bauthor{\bsnm{Hu}, \binits{Z.}},
\bauthor{\bsnm{Lu}, \binits{G.}},
\bauthor{\bsnm{Xu}, \binits{D.}}:
\bctitle{Fvc: A new framework towards deep video compression in feature space}.
In: \bbtitle{Proceedings of the IEEE Conference on Computer Vision and Pattern
  Recognition (CVPR)},
pp. \bfpage{1502}--\blpage{1511}
(\byear{2021}).
\doiurl{10.1109/CVPR46437.2021.00155}
\end{bchapter}
\endbibitem

\bibitem[\protect\citeauthoryear{Liu et~al.}{2021}]{dvc3}
\begin{barticle}
\bauthor{\bsnm{Liu}, \binits{H.}},
\bauthor{\bsnm{Lu}, \binits{M.}},
\bauthor{\bsnm{Ma}, \binits{Z.}},
\bauthor{\bsnm{Wang}, \binits{F.}},
\bauthor{\bsnm{Xie}, \binits{Z.}},
\bauthor{\bsnm{Cao}, \binits{X.}},
\bauthor{\bsnm{Wang}, \binits{Y.}}:
\batitle{Neural video coding using multiscale motion compensation and
  spatiotemporal context model}.
\bjtitle{IEEE Transactions on Circuits and Systems for Video Technology}
\bvolume{31}(\bissue{8}),
\bfpage{3182}--\blpage{3196}
(\byear{2021})
\doiurl{10.1109/TCSVT.2020.3035680}
\end{barticle}
\endbibitem

\bibitem[\protect\citeauthoryear{Lin et~al.}{2023}]{dvc4}
\begin{barticle}
\bauthor{\bsnm{Lin}, \binits{K.}},
\bauthor{\bsnm{Jia}, \binits{C.}},
\bauthor{\bsnm{Zhang}, \binits{X.}},
\bauthor{\bsnm{Wang}, \binits{S.}},
\bauthor{\bsnm{Ma}, \binits{S.}},
\bauthor{\bsnm{Gao}, \binits{W.}}:
\batitle{Dmvc: Decomposed motion modeling for learned video compression}.
\bjtitle{IEEE Transactions on Circuits and Systems for Video Technology}
\bvolume{33}(\bissue{7}),
\bfpage{3502}--\blpage{3515}
(\byear{2023})
\doiurl{10.1109/TCSVT.2022.3233221}
\end{barticle}
\endbibitem

\bibitem[\protect\citeauthoryear{Huang et~al.}{2022}]{dvc5}
\begin{botherref}
\oauthor{\bsnm{Huang}, \binits{Z.}},
\oauthor{\bsnm{Jia}, \binits{C.}},
\oauthor{\bsnm{Wang}, \binits{S.}},
\oauthor{\bsnm{Ma}, \binits{S.}}:
Hmfvc: A human-machine friendly video compression scheme.
IEEE Transactions on Circuits and Systems for Video Technology,
1--1
(2022)
\doiurl{10.1109/TCSVT.2022.3207596}
\end{botherref}
\endbibitem

\bibitem[\protect\citeauthoryear{Li et~al.}{2022}]{DCVC-HEM}
\begin{bchapter}
\bauthor{\bsnm{Li}, \binits{J.}},
\bauthor{\bsnm{Li}, \binits{B.}},
\bauthor{\bsnm{Lu}, \binits{Y.}}:
\bctitle{Hybrid spatial-temporal entropy modelling for neural video
  compression}.
In: \bbtitle{Proceedings of the 30th ACM International Conference on
  Multimedia},
pp. \bfpage{1503}--\blpage{1511}
(\byear{2022}).
\doiurl{10.1145/3503161.3547845}
\end{bchapter}
\endbibitem

\bibitem[\protect\citeauthoryear{Sheng et~al.}{2022}]{DCVC-TCM}
\begin{barticle}
\bauthor{\bsnm{Sheng}, \binits{X.}},
\bauthor{\bsnm{Li}, \binits{J.}},
\bauthor{\bsnm{Li}, \binits{B.}},
\bauthor{\bsnm{Li}, \binits{L.}},
\bauthor{\bsnm{Liu}, \binits{D.}},
\bauthor{\bsnm{Lu}, \binits{Y.}}:
\batitle{Temporal context mining for learned video compression}.
\bjtitle{IEEE Transactions on Multimedia}
\bvolume{25},
\bfpage{7311}--\blpage{7322}
(\byear{2022})
\doiurl{10.1109/TMM.2022.3220421}
\end{barticle}
\endbibitem

\bibitem[\protect\citeauthoryear{Zhou et~al.}{2015}]{jsccv1}
\begin{barticle}
\bauthor{\bsnm{Zhou}, \binits{C.}},
\bauthor{\bsnm{Lin}, \binits{C.-W.}},
\bauthor{\bsnm{Zhang}, \binits{X.}},
\bauthor{\bsnm{Guo}, \binits{Z.}}:
\batitle{A novel jscc scheme for uep-based scalable video transmission over
  mimo systems}.
\bjtitle{IEEE Transactions on Circuits and Systems for Video Technology}
\bvolume{25}(\bissue{6}),
\bfpage{1002}--\blpage{1015}
(\byear{2015})
\doiurl{10.1109/TCSVT.2014.2364418}
\end{barticle}
\endbibitem

\bibitem[\protect\citeauthoryear{Zhang et~al.}{2002}]{jsccv2}
\begin{barticle}
\bauthor{\bsnm{Zhang}, \binits{Q.}},
\bauthor{\bsnm{Ji}, \binits{Z.}},
\bauthor{\bsnm{Zhu}, \binits{W.}},
\bauthor{\bsnm{Zhang}, \binits{Y.-Q.}}:
\batitle{Power-minimized bit allocation for video communication over wireless
  channels}.
\bjtitle{IEEE Transactions on Circuits and Systems for Video Technology}
\bvolume{12}(\bissue{6}),
\bfpage{398}--\blpage{410}
(\byear{2002})
\doiurl{10.1109/TCSVT.2002.800322}
\end{barticle}
\endbibitem

\bibitem[\protect\citeauthoryear{Ji et~al.}{2001}]{jsccv3}
\begin{bchapter}
\bauthor{\bsnm{Ji}, \binits{Z.}},
\bauthor{\bsnm{Zhang}, \binits{Q.}},
\bauthor{\bsnm{Zhu}, \binits{W.}},
\bauthor{\bsnm{Lu}, \binits{J.}},
\bauthor{\bsnm{Zhang}, \binits{Y.-Q.}}:
\bctitle{Joint power control and source-channel coding for video communication
  over wireless networks}.
In: \bbtitle{Proceedings of the IEEE 54th Vehicular Technology Conference (VTC
  Fall)},
vol. \bseriesno{3},
pp. \bfpage{1658}--\blpage{1662}
(\byear{2001}).
\doiurl{10.1109/VTC.2001.956481}
\end{bchapter}
\endbibitem

\bibitem[\protect\citeauthoryear{Pei and Modestino}{2002}]{jsccv4}
\begin{bchapter}
\bauthor{\bsnm{Pei}, \binits{Y.}},
\bauthor{\bsnm{Modestino}, \binits{J.W.}}:
\bctitle{H. 263+ packet video over wireless ip networks using rate-compatible
  punctured turbo (rcpt) codes with joint source-channel coding}.
In: \bbtitle{Proceedings of the International Conference on Image Processing
  (ICIP)},
vol. \bseriesno{1},
pp. \bfpage{1}--\blpage{4}
(\byear{2002}).
\doiurl{10.1109/ICIP.2002.1038080}
\end{bchapter}
\endbibitem

\bibitem[\protect\citeauthoryear{Zhang et~al.}{2001}]{jsccv6}
\begin{bchapter}
\bauthor{\bsnm{Zhang}, \binits{Q.}},
\bauthor{\bsnm{Zhu}, \binits{W.}},
\bauthor{\bsnm{Ji}, \binits{Z.}},
\bauthor{\bsnm{Zhang}, \binits{Y.-Q.}}:
\bctitle{A power-optimized joint source channel coding for scalable video
  streaming over wireless channel}.
In: \bbtitle{Proceedings of the IEEE International Symposium on Circuits and
  Systems (ISCAS)},
vol. \bseriesno{5},
pp. \bfpage{137}--\blpage{140}
(\byear{2001}).
\doiurl{10.1109/ISCAS.2001.922004}
\end{bchapter}
\endbibitem

\bibitem[\protect\citeauthoryear{Jakubczak and Katabi}{2010a}]{soft1}
\begin{bchapter}
\bauthor{\bsnm{Jakubczak}, \binits{S.}},
\bauthor{\bsnm{Katabi}, \binits{D.}}:
\bctitle{Softcast: One-size-fits-all wireless video}.
In: \bbtitle{Proceedings of the ACM SIGCOMM Conference},
pp. \bfpage{449}--\blpage{450}
(\byear{2010}).
\doiurl{10.1145/1851182.1851257}
\end{bchapter}
\endbibitem

\bibitem[\protect\citeauthoryear{Jakubczak and Katabi}{2010b}]{soft2}
\begin{bchapter}
\bauthor{\bsnm{Jakubczak}, \binits{S.}},
\bauthor{\bsnm{Katabi}, \binits{D.}}:
\bctitle{Softcast: Clean-slate scalable wireless video}.
In: \bbtitle{Proceedings of the 2010 ACM Workshop on Wireless of the Students,
  by the Students, for the Students},
pp. \bfpage{9}--\blpage{12}
(\byear{2010}).
\doiurl{10.1145/1860039.1860043}
\end{bchapter}
\endbibitem

\bibitem[\protect\citeauthoryear{Lan et~al.}{2018}]{soft3}
\begin{barticle}
\bauthor{\bsnm{Lan}, \binits{C.}},
\bauthor{\bsnm{Luo}, \binits{C.}},
\bauthor{\bsnm{Zeng}, \binits{W.}},
\bauthor{\bsnm{Wu}, \binits{F.}}:
\batitle{A practical hybrid digital-analog scheme for wireless video
  transmission}.
\bjtitle{IEEE Transactions on Circuits and Systems for Video Technology}
\bvolume{28}(\bissue{7}),
\bfpage{1634}--\blpage{1647}
(\byear{2018})
\doiurl{10.1109/TCSVT.2017.2671417}
\end{barticle}
\endbibitem

\bibitem[\protect\citeauthoryear{Gui et~al.}{2020}]{soft4}
\begin{barticle}
\bauthor{\bsnm{Gui}, \binits{Y.}},
\bauthor{\bsnm{Lu}, \binits{H.}},
\bauthor{\bsnm{Jiang}, \binits{X.}},
\bauthor{\bsnm{Wu}, \binits{F.}},
\bauthor{\bsnm{Chen}, \binits{C.W.}}:
\batitle{Compressed pseudo-analog transmission system for remote sensing images
  over bandwidth-constrained wireless channels}.
\bjtitle{IEEE Transactions on Circuits and Systems for Video Technology}
\bvolume{30}(\bissue{9}),
\bfpage{3181}--\blpage{3195}
(\byear{2020})
\doiurl{10.1109/TCSVT.2019.2935127}
\end{barticle}
\endbibitem

\bibitem[\protect\citeauthoryear{He et~al.}{2023}]{soft5}
\begin{barticle}
\bauthor{\bsnm{He}, \binits{C.}},
\bauthor{\bsnm{Zhu}, \binits{S.}},
\bauthor{\bsnm{Zeng}, \binits{B.}}:
\batitle{Noma-based uncoded video transmission with optimization of joint
  resource allocation}.
\bjtitle{IEEE Transactions on Circuits and Systems for Video Technology}
\bvolume{33}(\bissue{5}),
\bfpage{2439}--\blpage{2450}
(\byear{2023})
\doiurl{10.1109/TCSVT.2022.3223954}
\end{barticle}
\endbibitem

\bibitem[\protect\citeauthoryear{Tung and G{\"u}nd{\"u}z}{2022}]{deepwive}
\begin{barticle}
\bauthor{\bsnm{Tung}, \binits{T.-Y.}},
\bauthor{\bsnm{G{\"u}nd{\"u}z}, \binits{D.}}:
\batitle{Deepwive: Deep-learning-aided wireless video transmission}.
\bjtitle{IEEE Journal on Selected Areas in Communications}
\bvolume{40}(\bissue{9}),
\bfpage{2570}--\blpage{2583}
(\byear{2022})
\doiurl{10.1109/JSAC.2022.3191354}
\end{barticle}
\endbibitem

\bibitem[\protect\citeauthoryear{Xu et~al.}{2023a}]{jsccv8}
\begin{barticle}
\bauthor{\bsnm{Xu}, \binits{J.}},
\bauthor{\bsnm{Tung}, \binits{T.-Y.}},
\bauthor{\bsnm{Ai}, \binits{B.}},
\bauthor{\bsnm{Chen}, \binits{W.}},
\bauthor{\bsnm{Sun}, \binits{Y.}},
\bauthor{\bsnm{G{\"u}nd{\"u}z}, \binits{D.D.}}:
\batitle{Deep joint source-channel coding for semantic communications}.
\bjtitle{IEEE Communications Magazine}
\bvolume{61}(\bissue{11}),
\bfpage{42}--\blpage{48}
(\byear{2023})
\doiurl{10.1109/MCOM.004.2200819}
\end{barticle}
\endbibitem

\bibitem[\protect\citeauthoryear{Xu et~al.}{2023b}]{jsccv9}
\begin{barticle}
\bauthor{\bsnm{Xu}, \binits{J.}},
\bauthor{\bsnm{Ai}, \binits{B.}},
\bauthor{\bsnm{Chen}, \binits{W.}},
\bauthor{\bsnm{Wang}, \binits{N.}},
\bauthor{\bsnm{Rodrigues}, \binits{M.}}:
\batitle{Deep joint source-channel coding for image transmission with visual
  protection}.
\bjtitle{IEEE Transactions on Cognitive Communications and Networking}
\bvolume{61}(\bissue{11}),
\bfpage{42}--\blpage{48}
(\byear{2023})
\doiurl{10.1109/TCCN.2023.3306851}
\end{barticle}
\endbibitem

\bibitem[\protect\citeauthoryear{Xu et~al.}{2023c}]{jsccv10}
\begin{barticle}
\bauthor{\bsnm{Xu}, \binits{J.}},
\bauthor{\bsnm{Ai}, \binits{B.}},
\bauthor{\bsnm{Wang}, \binits{N.}},
\bauthor{\bsnm{Chen}, \binits{W.}}:
\batitle{Deep joint source-channel coding for csi feedback: An end-to-end
  approach}.
\bjtitle{IEEE Journal on Selected Areas in Communications}
\bvolume{41}(\bissue{1}),
\bfpage{260}--\blpage{273}
(\byear{2023})
\doiurl{10.1109/JSAC.2022.3221963}
\end{barticle}
\endbibitem

\bibitem[\protect\citeauthoryear{Ranjan and Black}{2017}]{optic}
\begin{bchapter}
\bauthor{\bsnm{Ranjan}, \binits{A.}},
\bauthor{\bsnm{Black}, \binits{M.J.}}:
\bctitle{Optical flow estimation using a spatial pyramid network}.
In: \bbtitle{Proceedings of the IEEE Conference on Computer Vision and Pattern
  Recognition (CVPR)},
pp. \bfpage{2720}--\blpage{2729}
(\byear{2017}).
\doiurl{10.1109/CVPR.2017.291}
\end{bchapter}
\endbibitem

\bibitem[\protect\citeauthoryear{He et~al.}{2016}]{resnet}
\begin{bchapter}
\bauthor{\bsnm{He}, \binits{K.}},
\bauthor{\bsnm{Zhang}, \binits{X.}},
\bauthor{\bsnm{Ren}, \binits{S.}},
\bauthor{\bsnm{Sun}, \binits{J.}}:
\bctitle{Deep residual learning for image recognition}.
In: \bbtitle{Proceedings of the IEEE Conference on Computer Vision and Pattern
  Recognition (CVPR)},
pp. \bfpage{770}--\blpage{778}
(\byear{2016}).
\doiurl{10.1109/CVPR.2016.90}
\end{bchapter}
\endbibitem

\bibitem[\protect\citeauthoryear{Ball{\'e} et~al.}{2015}]{GDN1}
\begin{bchapter}
\bauthor{\bsnm{Ball{\'e}}, \binits{J.}},
\bauthor{\bsnm{Laparra}, \binits{V.}},
\bauthor{\bsnm{Simoncelli}, \binits{E.P.}}:
\bctitle{Density modeling of images using a generalized normalization
  transformation}.
In: \bbtitle{International Conference on Learning Representations (ICLR)},
pp. \bfpage{1}--\blpage{14}
(\byear{2015})
\end{bchapter}
\endbibitem

\bibitem[\protect\citeauthoryear{Islam et~al.}{}]{GDN2}
\begin{botherref}
\oauthor{\bsnm{Islam}, \binits{K.}},
\oauthor{\bsnm{Dang}, \binits{L.M.}},
\oauthor{\bsnm{Lee}, \binits{S.}},
\oauthor{\bsnm{Moon}, \binits{H.}}:
Image compression with recurrent neural network and generalized divisive
  normalization.
In: Proceedings of the IEEE Conference on Computer Vision and Pattern
  Recognition (CVPR),
pp. 1875--1879.
\doiurl{10.1109/CVPRW53098.2021.00209}
\end{botherref}
\endbibitem

\bibitem[\protect\citeauthoryear{Shi et~al.}{2016}]{espcn}
\begin{botherref}
\oauthor{\bsnm{Shi}, \binits{W.}},
\oauthor{\bsnm{Caballero}, \binits{J.}},
\oauthor{\bsnm{Husz{\'a}r}, \binits{F.}},
\oauthor{\bsnm{Totz}, \binits{J.}},
\oauthor{\bsnm{Aitken}, \binits{A.P.}},
\oauthor{\bsnm{Bishop}, \binits{R.}},
\oauthor{\bsnm{Rueckert}, \binits{D.}},
\oauthor{\bsnm{Wang}, \binits{Z.}}:
Real-time single image and video super-resolution using an efficient sub-pixel
  convolutional neural network.
2016 IEEE Conference on Computer Vision and Pattern Recognition (CVPR),
1874--1883
(2016)
\end{botherref}
\endbibitem

\bibitem[\protect\citeauthoryear{Jang et~al.}{2017}]{Gumbel-Softmax}
\begin{bchapter}
\bauthor{\bsnm{Jang}, \binits{E.}},
\bauthor{\bsnm{Gu}, \binits{S.}},
\bauthor{\bsnm{Poole}, \binits{B.}}:
\bctitle{Categorical reparameterization with gumbel-softmax}.
In: \bbtitle{Proceedings of the International Conference Learning Represent
  (ICLR)},
pp. \bfpage{1}--\blpage{13}
(\byear{2017})
\end{bchapter}
\endbibitem

\bibitem[\protect\citeauthoryear{Ball{\'e} et~al.}{2018}]{balle2018}
\begin{bchapter}
\bauthor{\bsnm{Ball{\'e}}, \binits{J.}},
\bauthor{\bsnm{Minnen}, \binits{D.}},
\bauthor{\bsnm{Singh}, \binits{S.}},
\bauthor{\bsnm{Hwang}, \binits{S.J.}},
\bauthor{\bsnm{Johnston}, \binits{N.}}:
\bctitle{Variational image compression with a scale hyperprior}.
In: \bbtitle{Proceedings of the International Conference Learning Represent
  (ICLR)},
pp. \bfpage{1}--\blpage{15}
(\byear{2018})
\end{bchapter}
\endbibitem

\bibitem[\protect\citeauthoryear{Lin et~al.}{2020}]{m-lvc}
\begin{bchapter}
\bauthor{\bsnm{Lin}, \binits{J.}},
\bauthor{\bsnm{Liu}, \binits{D.}},
\bauthor{\bsnm{Li}, \binits{H.}},
\bauthor{\bsnm{Wu}, \binits{F.}}:
\bctitle{M-lvc: Multiple frames prediction for learned video compression}.
In: \bbtitle{Proceedings of the IEEE Conference on Computer Vision and Pattern
  Recognition (CVPR)},
pp. \bfpage{3543}--\blpage{3551}
(\byear{2020}).
\doiurl{10.1109/CVPR42600.2020.00360}
\end{bchapter}
\endbibitem

\bibitem[\protect\citeauthoryear{Mentzer et~al.}{2022}]{nvc-gan}
\begin{bchapter}
\bauthor{\bsnm{Mentzer}, \binits{F.}},
\bauthor{\bsnm{Agustsson}, \binits{E.}},
\bauthor{\bsnm{Ball{\'e}}, \binits{J.}},
\bauthor{\bsnm{Minnen}, \binits{D.}},
\bauthor{\bsnm{Johnston}, \binits{N.}},
\bauthor{\bsnm{Toderici}, \binits{G.}}:
\bctitle{Neural video compression using gans for detail synthesis and
  propagation}.
In: \bbtitle{Proceedings of the European Conference Computer Vission (ECCV)},
pp. \bfpage{562}--\blpage{578}
(\byear{2022}).
\doiurl{10.1007/978-3-031-19809-032}
\end{bchapter}
\endbibitem

\bibitem[\protect\citeauthoryear{Guo et~al.}{2023}]{nvc-cross}
\begin{barticle}
\bauthor{\bsnm{Guo}, \binits{Z.}},
\bauthor{\bsnm{Feng}, \binits{R.}},
\bauthor{\bsnm{Zhang}, \binits{Z.}},
\bauthor{\bsnm{Jin}, \binits{X.}},
\bauthor{\bsnm{Chen}, \binits{Z.}}:
\batitle{Learning cross-scale weighted prediction for efficient neural video
  compression}.
\bjtitle{IEEE Transactions on Image Processing}
\bvolume{32},
\bfpage{3567}--\blpage{3579}
(\byear{2023})
\doiurl{10.1109/TIP.2023.3287495}
\end{barticle}
\endbibitem

\bibitem[\protect\citeauthoryear{Deng et~al.}{2009}]{imagenet}
\begin{bchapter}
\bauthor{\bsnm{Deng}, \binits{J.}},
\bauthor{\bsnm{Dong}, \binits{W.}},
\bauthor{\bsnm{Socher}, \binits{R.}},
\bauthor{\bsnm{Li}, \binits{L.-J.}},
\bauthor{\bsnm{Li}, \binits{K.}},
\bauthor{\bsnm{Fei-Fei}, \binits{L.}}:
\bctitle{Imagenet: A large-scale hierarchical image database}.
In: \bbtitle{Proceedings of the IEEE Conference on Computer Vision and Pattern
  Recognition (CVPR)},
pp. \bfpage{248}--\blpage{255}
(\byear{2009}).
\doiurl{10.1109/CVPR.2009.5206848}
\end{bchapter}
\endbibitem

\bibitem[\protect\citeauthoryear{Xue et~al.}{2019}]{vimon}
\begin{barticle}
\bauthor{\bsnm{Xue}, \binits{T.}},
\bauthor{\bsnm{Chen}, \binits{B.}},
\bauthor{\bsnm{Wu}, \binits{J.}},
\bauthor{\bsnm{Wei}, \binits{D.}},
\bauthor{\bsnm{Freeman}, \binits{W.T.}}:
\batitle{Video enhancement with task-oriented flow}.
\bjtitle{International Journal of Computer Vision}
\bvolume{127},
\bfpage{1106}--\blpage{1125}
(\byear{2019})
\doiurl{10.1007/s11263-018-01144-2}
\end{barticle}
\endbibitem

\bibitem[\protect\citeauthoryear{Bossen}{2013}]{hevc}
\begin{botherref}
\oauthor{\bsnm{Bossen}, \binits{F.}}:
``Common Test Conditions and Software Reference Configurations", Document
  JCTVC-L1100
\end{botherref}
\endbibitem

\bibitem[\protect\citeauthoryear{Wang et~al.}{2016}]{mcl}
\begin{bchapter}
\bauthor{\bsnm{Wang}, \binits{H.}},
\bauthor{\bsnm{Gan}, \binits{W.}},
\bauthor{\bsnm{Hu}, \binits{S.}},
\bauthor{\bsnm{Lin}, \binits{J.Y.}},
\bauthor{\bsnm{Jin}, \binits{L.}},
\bauthor{\bsnm{Song}, \binits{L.}},
\bauthor{\bsnm{Wang}, \binits{P.}},
\bauthor{\bsnm{Katsavounidis}, \binits{I.}},
\bauthor{\bsnm{Aaron}, \binits{A.}},
\bauthor{\bsnm{Kuo}, \binits{C.-C.J.}}:
\bctitle{Mcl-jcv: a jnd-based h. 264/avc video quality assessment dataset}.
In: \bbtitle{Proceedings of the IEEE International Conference on Image
  Processing (ICIP)},
pp. \bfpage{1509}--\blpage{1513}
(\byear{2016}).
\doiurl{10.1109/ICIP.2016.7532610}
\end{bchapter}
\endbibitem

\bibitem[\protect\citeauthoryear{Zhang et~al.}{2018}]{lpips}
\begin{bchapter}
\bauthor{\bsnm{Zhang}, \binits{R.}},
\bauthor{\bsnm{Isola}, \binits{P.}},
\bauthor{\bsnm{Efros}, \binits{A.A.}},
\bauthor{\bsnm{Shechtman}, \binits{E.}},
\bauthor{\bsnm{Wang}, \binits{O.}}:
\bctitle{The unreasonable effectiveness of deep features as a perceptual
  metric}.
In: \bbtitle{Proceedings of the IEEE Conference on Computer Vision and Pattern
  Recognition (CVPR)},
pp. \bfpage{586}--\blpage{595}
(\byear{2018})
\end{bchapter}
\endbibitem

\end{thebibliography}

\end{document}